\documentclass[11pt]{article}

\usepackage[margin=1in]{geometry}
\usepackage{microtype}
\usepackage{graphicx}
\usepackage{amsmath,amssymb}
\usepackage{siunitx}
\usepackage{authblk}
\usepackage{booktabs}
\usepackage{caption}
\usepackage{subcaption}
\usepackage{enumitem}
\usepackage{xcolor}
\usepackage{orcidlink}
\usepackage{hyperref}
\usepackage[nameinlink,capitalize]{cleveref}

\usepackage[
    backend=biber,
    style=numeric-comp,
    sorting=none,
    giveninits=true,
    maxbibnames=99
]{biblatex}

\setlist[itemize]{leftmargin=1.6em, itemsep=0.2em, topsep=0.2em}
\setlist[enumerate]{leftmargin=1.8em, itemsep=0.2em, topsep=0.2em}
\title{Bridging Atomistic Simulation and Experimental Processing Timescales with Goal-Directed Deep Reinforcement Learning}

\author[1]{Wonseok Jeong\thanks{Email: kensou444@gmail.com}\,\orcidlink{0000-0001-8894-1857}}
\author[1]{Francesca Tavazza\,\orcidlink{0000-0002-5602-180X}}
\author[1]{Brian DeCost\thanks{Email: brian.decost@nist.gov}\,\orcidlink{0000-0002-3459-5888}}
\affil[1]{Material Measurement Laboratory, National Institute of Standards and Technology (NIST), Gaithersburg, MD, USA}

\begin{document}
\maketitle

\begin{abstract}

Atomic-scale modeling capabilities have advanced rapidly through integration of machine learning, yet a key bottleneck remains. Even with an accurate potential energy surface and a clear target material, we still lack a practical atomistic dynamics framework that can simulate how materials form under realistic synthesis and processing conditions. Many processing transformations are governed by rare events in non-idealized evolving environments, while direct molecular dynamics is limited to femtosecond timesteps and nanosecond to microsecond trajectories. Existing acceleration methods often require prior mechanistic knowledge such as reaction coordinates, collective variables, event tables, or pathway guesses, which is rarely available in real experiments. Here we present an $E(3)$-equivariant deep reinforcement learning framework that enables goal-directed pathway discovery without hand-crafted reaction coordinates. The framework introduces a complementary operating mode for atomistic simulation in which realistic, non-idealized environments can be addressed directly, without predefined reaction coordinates or idealized mechanism assumptions. As a challenging benchmark, we target silicon dry oxidation, where rare-event pathways in amorphous SiO$_2$ are effectively inaccessible to conventional atomistic methods. We treat an O$_2$ molecule as an agent that performs continuous rigid-body translations and rotations in a Si/a-SiO$_2$ environment. The agent is trained with an episode-level objective that rewards verified O$_2$ dissociation while preferring low effective activation barriers. We demonstrate that the learned policy discovers kinetically favorable O$_2$ diffusion and dissociation pathways in a disordered Si/a-SiO$_2$ environment, progressively improving success rate while reducing effective activation barriers over training. We also discuss how the approach can be generalized to other processing and synthesis problems.

\end{abstract}

\section{Introduction}
Over the past decade, the materials community has made rapid progress by integrating machine learning (ML) into atomistic simulation, including property prediction from atomic structure, machine-learning interatomic potentials (MLIPs) that reproduce first-principles-level energies and forces at scale, and inverse design workflows that propose candidate compositions and microstructures with targeted properties.~\cite{behler2007generalized,schutt2017schnet,xie2018crystal,merchant2023scaling,wood2025family,liao2023equiformerv2,batzner20223,musaelian2023learning,zeni2025generative,xie2021crystal,yang2024mattersim,deng2023chgnet,batatia2022mace,choudhary2021atomistic,kang2022accelerated} Together, these developments have reshaped how we model materials and how we identify promising candidates. However, a separate and increasingly consequential bottleneck remains largely overlooked in the current ML-for-materials landscape. Even if we can design a material and evaluate it with an accurate potential energy surface, we still lack a practical framework to simulate how materials actually evolve under experimentally relevant synthesis and processing conditions. More broadly, this reflects a long-recognized limitation of atomistic simulation: although atomic-scale fidelity is available, the timescales directly accessible to conventional methods remain far shorter than those relevant to many experimental processes.~\cite{uberuaga2020computational} As a result, the field has become increasingly capable at identifying stable materials with favorable properties, yet our capability remains limited in explaining how systems actually reach those states when the pathway is complex, heterogeneous, and extended in time. We refer to this mismatch between predictive capability and process-realization capability as a realization gap in atomistic simulation.

The realization gap matters because many technologically important materials phenomena are pathway-determined at the atomic scale. Microstructure, defect populations, interfacial chemistry, and metastable phases are often not uniquely specified by equilibrium thermodynamics.~\cite{ashby2018chapter19,Rudolph2016} In practice, realized materials outcomes are governed not only by the material itself, but also by the process by which it is formed. They depend on history, kinetics, and local environments that are spatially disordered and temporally evolving. Processing rarely resembles an idealized single-mechanism event on a clean surface or a crystalline lattice with a known reaction coordinate. Instead, real materials processing proceeds in messy, dynamic settings such as amorphous networks with broad distributions of local geometries, evolving interfaces, competing reaction channels, variable local stresses and chemical concentrations, and heterogeneous defect landscapes. Bridging atomistic modeling to these realities is essential for connecting computational predictions to manufacturing outcomes, interpreting \textit{in situ} characterization, and establishing mechanistic control over processing windows rather than relying on post hoc explanations.

Why does this realization gap persist despite increasingly accurate interatomic potentials? The central reason is that the standard atomistic dynamics toolbox remains fundamentally constrained by timescale, prior-knowledge requirements, and search efficiency. Molecular dynamics (MD) integrates Newton’s equations of motion and is physically transparent, but it must resolve atomic vibrations with femtosecond timesteps.~\cite{frenkel2023understanding} Even microsecond trajectories therefore require billions of steps, whereas many processing events occur on far longer timescales. This limitation is not removed simply by using larger computers, because conventional MD is intrinsically serial in time: one timestep must be completed before the next begins, so parallelization primarily extends system size rather than temporal reach. In rare-event-dominated systems, the dynamics consist of long waiting periods dominated by local vibrational motion, punctuated by relatively rapid activated transitions between metastable configurations.~\cite{uberuaga2020computational} Direct MD thus spends most of its computational effort resolving vibrations within basins rather than the transitions that control chemistry and microstructural evolution. Raising the temperature can increase event frequency within an accessible MD window, but it often alters the operative mechanism and compromises kinetic fidelity.~\cite{so2000temperature} For this reason, brute-force MD is not a practical route to processing-relevant trajectories, even with an accurate potential.~\cite{uberuaga2020computational}

Methods designed to accelerate rare events can substantially reduce the sampling burden, but many are most effective when the process of interest is already sufficiently characterized. In broad terms, long-timescale rare-event methods can be viewed as kinetic or dynamic approaches: kinetic methods are especially powerful when one can identify relevant initial and final regions, enumerate plausible events, or construct an informative reaction coordinate, whereas truly dynamical methods are needed when such prior knowledge is not available.~\cite{uberuaga2020computational} Within this broader landscape, approaches such as nudged elastic band (NEB), kinetic Monte Carlo (kMC), and metadynamics can be highly effective when an appropriate pathway guess, event catalog, or set of collective variables is available.~\cite{henkelman2000climbing,voter2007introduction,laio2002escaping} Accordingly, automatically learning informative collective variables has itself become an active research direction.~\cite{Bhakat2022} In realistic processing environments, however, non-ideal structures and evolving interfaces often support many competing micro-pathways, and the relevant descriptors are rarely obvious in advance. When the assumed mechanism is incomplete or incorrect, these methods can spend substantial effort exploring irrelevant degrees of freedom rather than the transformations that actually control the process.

Potential energy surface exploration methods such as the activation--relaxation technique (ART) and the dimer method search more directly for saddle points, but they can still become inefficient in rugged, high-dimensional landscapes where productive directions are sparse, state-dependent, and difficult to identify reliably.~\cite{barkema1996event,mousseau1998traveling,henkelman1999dimer} In such settings, curvature can be noisy, searches can drift into unproductive directions, and repeated restarts may be required. More broadly, traditional atomistic simulation is not intrinsically goal-directed. These approaches follow physical dynamics or local saddle-point structure, but they do not explicitly aim to accomplish a specified chemically meaningful transformation within a limited computational budget. For many processing problems, however, the practical objective is not generic exploration but the discovery of a conditional trajectory that achieves an event such as interfacial reaction, phase transformation, or transport through a disordered network while remaining kinetically plausible. Without such guidance, substantial computation can be spent on physically allowed yet chemically unproductive exploration. Generative approaches such as diffusion models can also aim to model materials transformations, but they learn a static configuration distribution from the training set rather than a kinetics-consistent pathway. Because training data typically consist of equilibrium or near-equilibrium structures, the intermediate denoising states are not expected to correspond to physical transition intermediates.~\cite{ho2020denoising,kwon2024spectroscopy} More fundamentally, learning to generate rare pathway states would still require examples of those rare intermediates in the first place, which faces the same sampling bottleneck as direct simulation.

These limitations together create a fundamental obstacle for simulating synthesis and processing as they occur in experiment.~\cite{uberuaga2020computational} Addressing the realization gap therefore requires a framework that can (i) discover productive pathways without hand-crafted reaction coordinates, (ii) steer exploration toward chemically meaningful transformations rather than generic low-level motion, and (iii) retain kinetic plausibility by preferring low-barrier transitions over unphysical shortcuts. Reinforcement learning (RL) offers a natural mechanism for this objective: an agent can interact with an atomistic environment, make the system evolve, and learn from rewards that encode processing-relevant goals and physical constraints.~\cite{sutton1998reinforcement}

In this work, we introduce REALIZE, a reinforcement-learning framework for atomistic long-timescale simulation in experimentally relevant environments. Rather than prescribing reaction coordinates or enumerating events in advance, REALIZE learns a symmetry-consistent control policy that steers trajectories toward a specified transformation while explicitly preferring low-barrier transitions. This defines a goal-directed, barrier-aware operating mode for atomistic simulation in disordered and evolving environments where reaction coordinates and event catalogs are difficult to define and process-relevant mechanisms remain inaccessible to existing approaches. 

\subsection{Case study: silicon dry oxidation as a processing-relevant benchmark}

We focus on silicon dry oxidation as a representative system in which the realization gap is both technologically important and scientifically well contextualized. Thermal oxidation of silicon forms the thin amorphous SiO$_2$ layer that underpins MOSFET operation. The resulting Si/a-SiO$_2$ interface, often part of modern gate stacks such as Si/a-SiO$_2$/HfO$_2$, strongly influences carrier transport, leakage current, and long-term device reliability.~\cite{wilk2001high,robertson2015high,liu2019ab} A notable advantage of this platform is that a relatively simple thermal oxidation process can produce exceptionally high-quality Si/a-SiO$_2$ interfaces that are clean, abrupt, and low in defect density.~\cite{green2001ultrathin,lai1999limiting,ross1992dynamic} Even so, residual interface defects can act as carrier traps and degrade device performance.~\cite{lenahan1998can} As device dimensions shrink and architectures move toward FinFET and gate-all-around designs, device behavior becomes increasingly sensitive to local interfacial structure and even individual defects, strengthening the need for atomic-scale mechanistic understanding.~\cite{hisamoto2000finfet,loubet2017stacked}

Silicon oxidation has been studied extensively for more than half a century and remains one of the foundational processes in semiconductor technology.~\cite{deal1965general,ghez1972kinetics,hopper1975thermal,massoud1985thermal} The kinetics of thermal oxidation are classically described by the Deal--Grove model, which separates the process into an initial rapid surface reaction followed by a diffusion-limited regime.~\cite{deal1965general,ghez1972kinetics,green2001ultrathin,cvitkovich2023dynamic} In dry oxidation, O$_2$ is the oxidant, whereas in wet oxidation H$_2$O plays this role.~\cite{bakos2002reactions} After the early stage, the rate-limiting step is widely accepted to be the diffusion of O$_2$ through amorphous SiO$_2$ toward the Si/a-SiO$_2$ interface, where oxidation occurs. Isotope experiments have shown that the O$_2$ molecule remains intact during diffusion and reacts at the interface without measurable exchange with the SiO$_2$ network, consistent with oxidation controlled primarily by oxidant transport through the oxide rather than by Si transport.~\cite{rosencher197918o,rochet198418o} Because this system is so well characterized experimentally, it provides a valuable benchmark for validating a new atomistic simulation framework against both established macroscopic behavior and known barrier scales.

Despite this extensive body of work, a clear and complete picture of the atomic-scale dynamics remains elusive. The amorphous SiO$_2$ network presents a highly disordered free-volume landscape with many competing micro-pathways. Reported characteristic barriers for oxygen transport are on the order of \SI{1.2}{eV}.~\cite{lamkin1992oxygen,bongiorno2004multiscale} At a typical dry-oxidation processing temperature of \SI{1000}{K}, such a barrier corresponds to characteristic waiting times on the order of tens of nanoseconds, whereas MD operates with femtosecond timesteps. Even when an event is kinetically accessible, direct MD spends millions of steps dominated by vibrational motion before observing a single relevant transition. The multiplicity of pathways in the amorphous network also makes it difficult to define a compact set of collective variables for enhanced sampling or to construct a complete event catalog for kMC. Similarly, systematic saddle-point searches in a rugged potential energy surface with many shallow traps can drift into irrelevant directions or require extensive restarts.

These features make silicon dry oxidation an ideal benchmark for a goal-directed, symmetry-preserving reinforcement learning framework. It is technologically critical, experimentally well studied, and kinetically controlled by rare events in a disordered environment. Demonstrating that a model can discover low-barrier diffusion pathways and interface reactions in this setting therefore provides a meaningful test of its ability to bridge atomistic dynamics with realistic processing conditions.

\subsection{Prior works and design rationale}

Recent reinforcement learning studies in atomistic simulation have demonstrated encouraging results for rare-event discovery, transition-state identification, inverse design, and long-timescale transport.~\cite{lan2021discovering,nomura2024molecular,lan2024enabling,tang2024reinforcement,bihani2023stridernet,banik2021learning,banik2023continuous,yoon2021deep} However, many existing formulations still rely on problem-specific state and action abstractions that limit physical generality. Depending on the application, these abstractions may take the form of discretized spatial representations, predefined reaction events, local perturbation operators, graph-based displacement rules, or search moves over structural parameters. Such designs can be highly effective within their intended domains, but they often tie the policy to a particular representation, search procedure, or system setup rather than providing a more direct and generally transferable formulation of atomistic transformation. In addition, many prior formulations optimize local transitions, endpoint stability, or search efficiency more directly than the process-level progression of a targeted transformation. As a result, a substantial fraction of computation can still be spent on physically allowed yet chemically unproductive exploration.

Our design rationale is to move from representation-specific search toward a physically structured control policy for atomistic transformation. In the setting of disordered processing, the relevant challenge is not merely to locate a low-energy state or a plausible local event, but to discover a sequence of kinetically favorable transitions that advances a specified chemical process in an evolving environment. This motivates a framework that combines goal-directed exploration, explicit barrier preference, and symmetry-consistent action generation within a single closed-loop atomistic simulation framework.

\section{RL formulation}
\begin{figure}[htbp]
    \centering
      \includegraphics[width=\textwidth]{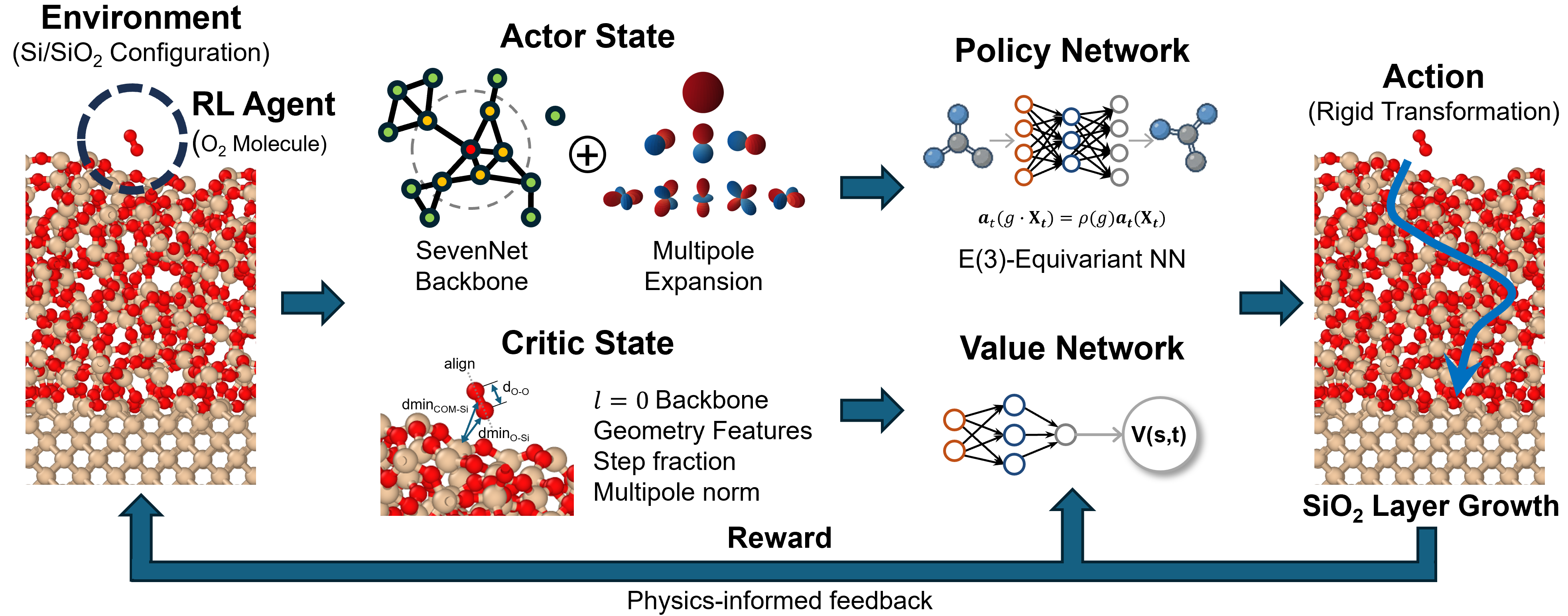}
\caption{
Overview of REALIZE, the $E(3)$-equivariant deep reinforcement learning framework. The environment is a Si/$a$-SiO$_2$ configuration containing an O$_2$ agent. At each step, the current atomistic configuration is encoded into actor and critic representations using a SevenNet backbone together with an environmental multipole. The policy network proposes a rigid-body update of the O$_2$ molecule, parameterized by translation direction, translation magnitude, and bond-axis orientation. The proposed rigid-body move is applied, after which the rigid-body constraint is released and the structure is relaxed on a reactive ReaxFF potential energy surface. The resulting transition is evaluated through dissociation checks and NEB-based barrier estimation. Rewards combine verified chemical progress with barrier preference, and the policy and value networks are optimized in closed loop. Successful oxidation events are retained in the evolving environment, so the oxide structure progressively changes during training.
}
\label{fig:overview}
\end{figure}

\subsection{Overview}

REALIZE is a symmetry-preserving continuous-control framework for atomistic rare-event dynamics in disordered and evolving environments (Fig.~\ref{fig:overview}). The central idea is to combine goal-directed exploration with explicit kinetic preference while enforcing $E(3)$ symmetry so that the learned behavior is not tied to an arbitrary simulation frame. The agent is an O$_2$ molecule, and each action is a continuous rigid-body move. Directional outputs transform equivariantly under global rotations and translations, while scalar quantities such as translation magnitude and value prediction remain invariant.

At each step, the current Si/a-SiO$_2$ structure together with the O$_2$ configuration is encoded into an equivariant state using a SevenNet MLIP backbone~\cite{park2024scalable} augmented by a long-range environmental multipole centered at the O$_2$ center of mass. From this state, the policy proposes a rigid-body update for the O$_2$ molecule by first selecting a translation direction on $S^2$ and then predicting the associated translation magnitude and molecular orientation. This rigid-body constraint applies only to the policy-proposed displacement. After the move is applied, all chemically active atoms are relaxed on a reactive ReaxFF potential energy surface, so the O--O bond is free to stretch, react with the surrounding network, or dissociate if the proposed configuration lies near a reactive pathway. The resulting transition is then evaluated. Kinetic plausibility is assessed using NEB-based barrier estimates, while dissociation-verification and rollback rules reject unstable or nonphysical transitions.

This interaction defines one step of the RL environment. The process is repeated until the episode terminates. An episode ends either when O$_2$ dissociation is verified, corresponding to a successful oxidation event, or when a termination condition is reached, such as failure to dissociate within the maximum number of steps, escape of O$_2$ from the active region, or repeated catastrophic NEB failures. Rewards combine verified chemical progress with barrier preference, and the policy and value networks are optimized using Proximal Policy Optimization (PPO).~\cite{schulman2017proximal}

Beyond a single episode, the framework operates in an evolving environment rather than resetting after every successful oxidation event. When an O$_2$ molecule dissociates, the oxidized structure is retained, so subsequent episodes continue from the updated Si/a-SiO$_2$ environment. In this way, the local interface and oxide network progressively evolve during training. We refer to the sequence of episodes carried out before resetting the environment as one cycle. After each cycle, the reference barrier level in the reward function is decreased according to a fixed schedule, so that the optimization criterion becomes progressively more stringent over training

Our spherical-harmonics directional parameterization is conceptually inspired by symmetry-equivariant generative models such as SYMPHONY,~\cite{daigavane2023symphony} but the setting here is fundamentally different: the policy is learned in a closed-loop physics environment and optimized for long-horizon, barrier-aware atomistic dynamics rather than unconditional structure generation.

\subsection{Environment preparation}

The environment is an atomistic Si/a-SiO$_2$ configuration in Cartesian coordinates. It consists of a crystalline Si substrate and an amorphous surface oxide layer generated by explicit reactive MD rather than by manual construction. The starting substrate is a Si(001) slab containing 384 Si atoms in a simulation cell of dimensions $21.45 \times 21.45 \times 44.75$~\AA$^3$, with periodic boundary conditions applied in the lateral directions. To prepare the initial oxide, we performed reactive O$_2$ deposition simulations at \SI{1000}{K} using the LAMMPS code package with a ReaxFF potential.~\cite{thompson2022lammps,noaki2023development} In these simulations, O$_2$ molecules were inserted into a deposition region above the surface and evolved dynamically under NVE integration, while the substrate atoms were thermostatted with NVT and the bottom part of the Si slab was held fixed to mimic a bulk-like support. This setup captures the early oxidation regime, where oxidation proceeds near the exposed Si surface through relatively accessible local reactions and does not yet require activated diffusion through a pre-existing amorphous oxide. After approximately \SI{20}{ns} of deposition dynamics, the oxide thickness converged to about \SI{5}{\angstrom}. Continued simulation up to approximately \SI{70}{ns} showed no meaningful O$_2$ dissociation events beyond this initial growth stage, indicating that the rapid initial oxidation regime had been exhausted and that subsequent oxidation would require diffusion through the formed amorphous SiO$_2$ layer. We therefore use the resulting Si/a-SiO$_2$ structure as the starting environment for reinforcement learning, where the rare-event diffusion and dissociation processes become the central focus.

In this work, the potential energy surface is described using ReaxFF to enable efficient relaxation and barrier evaluations with reasonable accuracy.~\cite{noaki2023development} All MD simulations in this work, including the initial reactive deposition simulations used to prepare the Si/a-SiO$_2$ environment, the potential-energy-surface evaluations and structural relaxations during reinforcement learning, NEB calculations and the post-step verification MD, were carried out using the LAMMPS code package.~\cite{henkelman2000climbing,thompson2022lammps} The REALIZE framework itself is not tied to ReaxFF and can, in principle, be coupled to other potential energy surface models, including MLIP or first-principles calculations.~\cite{cvitkovich2024machine}

\subsection{Agent}

The agent is an O$_2$ molecule represented by the Cartesian positions of its two oxygen atoms. We model a single O$_2$ agent per oxidation event because even at a typical dry-oxidation pressure of 1~atm the expected gas-phase occupancy in our simulation cell is well below one (approximately $0.15$ O$_2$ molecules in the given simulation-cell volume at $1$~atm and $1000$~K from an ideal-gas estimate), so direct O$_2$--O$_2$ interactions can be neglected. Moreover, the region of primary interest is within and near the oxide, where the effective O$_2$ activity is lower than in the external gas phase.

At each step $t$, the agent observes the current atomistic configuration $\mathbf{X}_t$ and proposes an action $\mathbf{a}_t$ corresponding to a rigid-body move of the O$_2$ molecule. The environment applies the proposed move, relaxes the structure to a nearby local minimum, and evaluates the proposed transition. The environment then returns the next configuration $\mathbf{X}_{t+1}$, a termination flag, and bookkeeping quantities used for learning and diagnostics. An episode is a finite sequence of such steps. It terminates when O$_2$ dissociation is verified or when a termination condition is met, such as reaching the maximum step count or rejecting an unstable transition. Rewards are assigned at the end of the episode based on the overall outcome and a barrier-based figure of merit computed from the stepwise transition barriers encountered along the trajectory.

\subsection{Actions}
At each step, the agent proposes a rigid-body action consisting of a translation direction, a translation magnitude, and a molecular reorientation. We write the action at step $t$ as
\begin{equation}
\mathbf{a}_t = \bigl(\hat{\mathbf{d}}_t,\, s_t,\, \hat{\mathbf{b}}_t\bigr),
\end{equation}
where $\hat{\mathbf{d}}_t \in S^2$ is the unit translation direction, $s_t \in [s_{\min}, s_{\max}]$ is the translation magnitude, and $\hat{\mathbf{b}}_t \in S^2$ is the directed O--O bond axis, defined as the unit vector from atom O($a$) to atom O($b$). Here O($a$) and O($b$) denote the two oxygen atoms in the O$_2$ molecule, with fixed atom labels preserved throughout an episode. The corresponding translation displacement is
\begin{equation}
\Delta \mathbf{x}_t = s_t \hat{\mathbf{d}}_t.
\end{equation}

Because O$_2$ has $D_{\infty h}$ symmetry, its orientation is fully specified by its bond axis, so we predict $\hat{\mathbf{b}}_t$ rather than a full $3\times 3$ rotation matrix. An undirected axis would identify $\hat{\mathbf{b}}_t$ and $-\hat{\mathbf{b}}_t$, but NEB interpolation preserves atom labels and can therefore produce different interpolated pathways and different apparent barriers when the bond direction is flipped. For this reason, we model the bond axis as a directed unit vector.

\subsection{\texorpdfstring{$E(3)$}{E(3)} equivariance and invariances}
We enforce $E(3)$ symmetry end to end so that the learned policy is not tied to an arbitrary simulation frame. Let $\mathbf{X}_t$ denote the atomic configuration at step $t$, and let
\begin{equation}
\mathbf{a}(\mathbf{X}_t)
=
\bigl(\hat{\mathbf{d}}(\mathbf{X}_t),\, s(\mathbf{X}_t),\, \hat{\mathbf{b}}(\mathbf{X}_t)\bigr)
\end{equation}
denote the action predicted from that configuration. Then, for any global rigid transformation $g \in E(3)$,
\begin{equation}
\mathbf{a}(g \cdot \mathbf{X}_t) = \rho(g)\,\mathbf{a}(\mathbf{X}_t),
\end{equation}
where $\rho(g)$ denotes the induced action on the mixed action space. Since the action contains two directional components and one scalar component, $\rho(g)$ rotates the directional components and leaves the scalar component invariant:
\begin{equation}
\rho(g)(\hat{\mathbf{d}}, s, \hat{\mathbf{b}})
=
(R\hat{\mathbf{d}},\, s,\, R\hat{\mathbf{b}}),
\end{equation}
with $R \in SO(3)$ the rotational part of $g$.

Equivalently, the directional outputs transform equivariantly,
\begin{equation}
\hat{\mathbf{d}}(g \cdot \mathbf{X}_t) = R\,\hat{\mathbf{d}}(\mathbf{X}_t),
\qquad
\hat{\mathbf{b}}(g \cdot \mathbf{X}_t) = R\,\hat{\mathbf{b}}(\mathbf{X}_t),
\end{equation}
whereas the scalar outputs are invariant,
\begin{equation}
s(g \cdot \mathbf{X}_t) = s(\mathbf{X}_t),
\qquad
V(g \cdot \mathbf{X}_t) = V(\mathbf{X}_t).
\end{equation}

The distinction between $SE(3)$ and the full $E(3)$ lies in the treatment of inversion. In the present work we follow the SevenNet convention of using even parity throughout, so the realized symmetry is effectively $SE(3)$ rather than full $O(3)$ equivariance. Using both even- and odd-parity irreducible representations would allow the same framework to represent the full $E(3)$ action.

\subsection{State Representation}
\begin{figure}[htbp]
    \centering
    \includegraphics[width=12cm]{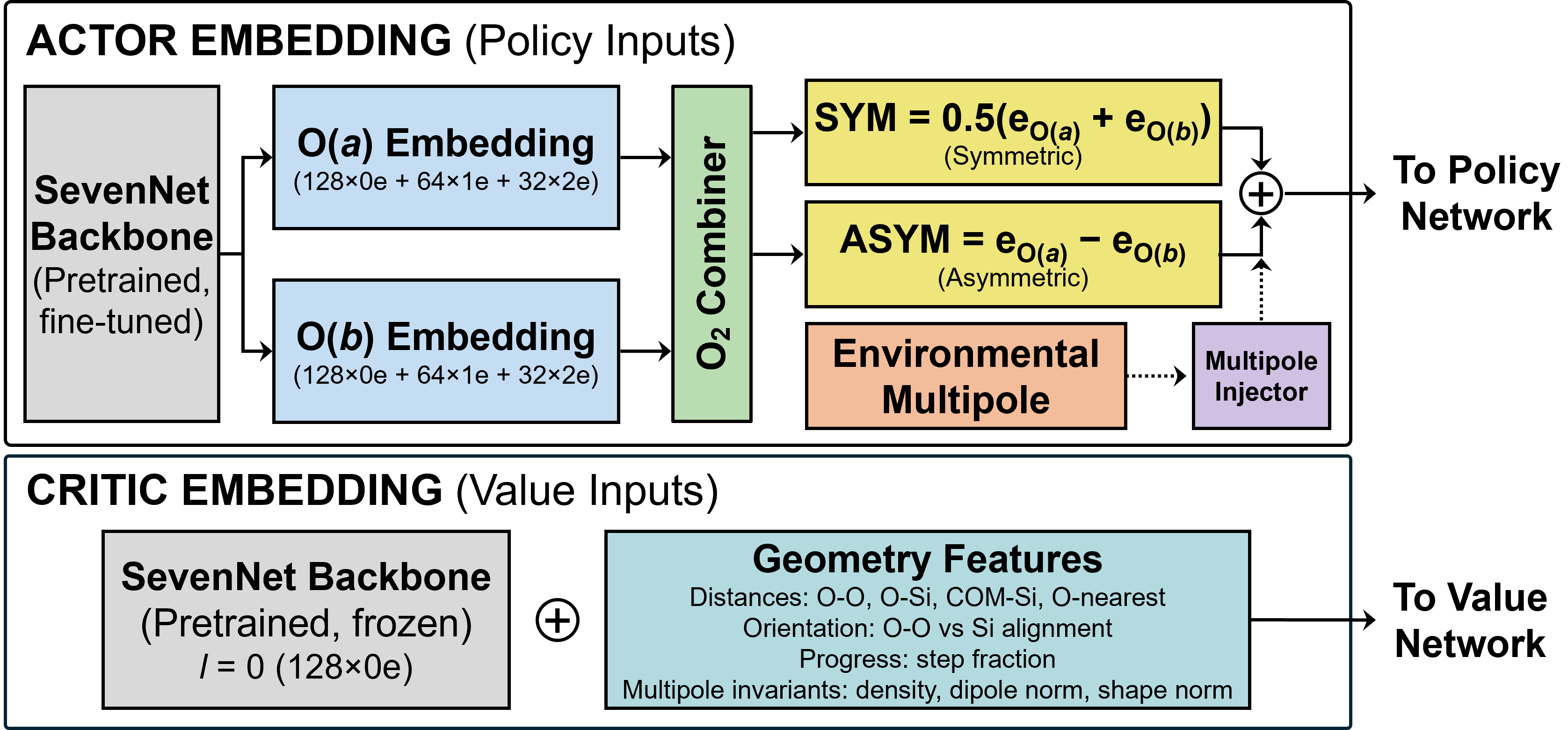}
\caption{
State representation used by the actor and critic. For the actor, O-centered equivariant embeddings from the SevenNet backbone are combined into a symmetric channel, $\mathrm{SYM} = \tfrac{1}{2}(e_{\mathrm{O}(a)} + e_{\mathrm{O}(b)})$, and an antisymmetric channel, $\mathrm{ASYM} = e_{\mathrm{O}(a)} - e_{\mathrm{O}(b)}$. The symmetric channel is augmented with an environmental multipole through an equivariant injector and is used as input to the policy network. The antisymmetric channel is used by the O--O axis head to retain atom-label information. For the critic, only the scalar ($\ell=0$) backbone block is used together with hand-crafted geometric invariants, including distances, orientation measures, step fraction, and multipole invariants. This separation lets the actor use a rich equivariant representation while keeping the critic input strictly invariant.
}
    \label{fig:state}
\end{figure}
\subsubsection{Pretrained SevenNet backbone}
Per-atom $E(3)$-equivariant embeddings are extracted using a SevenNet graph interaction network. The SevenNet MLIP is pretrained against a ReaxFF reference potential energy surface.~\cite{park2024scalable} The training set consists primarily of MD snapshots from the early stage of silicon oxidation obtained during preparation of the initial Si/a-SiO$_2$ environment. To broaden the local chemical and configurational coverage relevant to the present task, we additionally include snapshots in which O$_2$ molecules are inserted into Si/a-SiO$_2$ structures, as well as trajectories describing O$_2$ diffusion within amorphous SiO$_2$. Both total energies and atomic forces are used to train the MLIP.

For each oxygen atom, the pretrained MLIP provides an equivariant embedding with irreducible representations
\begin{equation}
128 \times 0e \;+\; 64 \times 1e \;+\; 32 \times 2e,
\end{equation}
corresponding to 128 scalars ($\ell=0$), 64 vectors ($\ell=1$), and 32 $\ell=2$ features, for a total of 480 dimensions per oxygen atom.

Two embeddings are produced, centered on O($a$) and O($b$), and are combined without additional learnable parameters:
\begin{align}
\mathrm{SYM} &= \tfrac{1}{2}\left(\mathbf{e}_{\mathrm{O}(a)} + \mathbf{e}_{\mathrm{O}(b)}\right), \\
\mathrm{ASYM} &= \mathbf{e}_{\mathrm{O}(a)} - \mathbf{e}_{\mathrm{O}(b)}.
\end{align}
SYM is swap-invariant and is used for translation-direction and translation-magnitude prediction, while ASYM is swap-antisymmetric and is used for prediction of the directed bond orientation.

During RL training, the last two interaction layers of the backbone are selectively unfrozen and fine-tuned at a reduced learning rate. Because the backbone was pretrained for scalar energy prediction, its $\ell \ge 1$ channels initially encode chemically relevant directional information in a vector/tensor basis but are not directly optimized for the transport-direction task. Fine-tuning allows these channels to adapt through RL gradients, producing environment-dependent directional features that complement the environmental multipole described below. To prevent this adaptation from destabilizing value learning, we maintain a separate frozen copy of the backbone for the critic, so that the value function always sees a stationary representation.

\subsubsection{Environmental multipole embedding}
Although the backbone embedding captures the local chemical environment within its interaction cutoff, a complementary long-range directional signal accelerates early learning. We construct an environmental multipole centered at the O$_2$ center of mass:
\begin{equation}
c_{\ell m} = \sum_{j} w_{Z_j}\, f(r_j)\, Y_{\ell m}(\hat{\mathbf{r}}_j),
\end{equation}
where $\mathbf{r}_j = \mathbf{R}_j - \mathbf{R}_{\mathrm{COM}}$ is the position of substrate atom $j$ relative to the O$_2$ center of mass, $\hat{\mathbf{r}}_j = \mathbf{r}_j/\|\mathbf{r}_j\|$, $Y_{\ell m}$ denotes the real spherical harmonics, $w_{Z_j}$ is a learnable species-dependent weight, and
\begin{equation}
f(r_j) = \tfrac{1}{2}\left(1 + \cos\left(\pi r_j / r_{\mathrm{cut}}\right)\right)
\end{equation}
is a cosine cutoff with $r_{\mathrm{cut}} = \SI{9.0}{\angstrom}$, providing a smooth radial envelope.~\cite{behler2011atom} We retain multipoles up to $\ell=2$, giving the irreducible representation
\begin{equation}
1\times 0e + 1\times 1e + 1\times 2e.
\end{equation}
The $\ell=0$ term represents the surrounding weighted density, the $\ell=1$ term provides a dipolar directional signal, and the $\ell=2$ term captures quadrupolar anisotropy of the nearby environment. This construction remains $E(3)$-equivariant under global rigid transformations.

The multipole coefficients are mapped into the same irreducible-representation structure as SYM through an equivariant linear layer and added to the backbone embedding used by the translation-direction head. Early in training, before the backbone's directional channels have adapted, the multipole provides the dominant long-range directional signal. As training progresses and the backbone fine-tunes, its $\ell \ge 1$ features increasingly carry environment-specific directional information that the global multipole cannot resolve, such as the locations of nearby insertion channels. The O--O axis head uses ASYM, whereas the critic and translation-magnitude head operate on invariant scalar features derived after conditioning on the sampled direction.

\subsection{Policy network (actor)}
\subsubsection{Autoregressive factorization}
\begin{figure}[htbp]
    \centering
      \includegraphics[width=\textwidth]{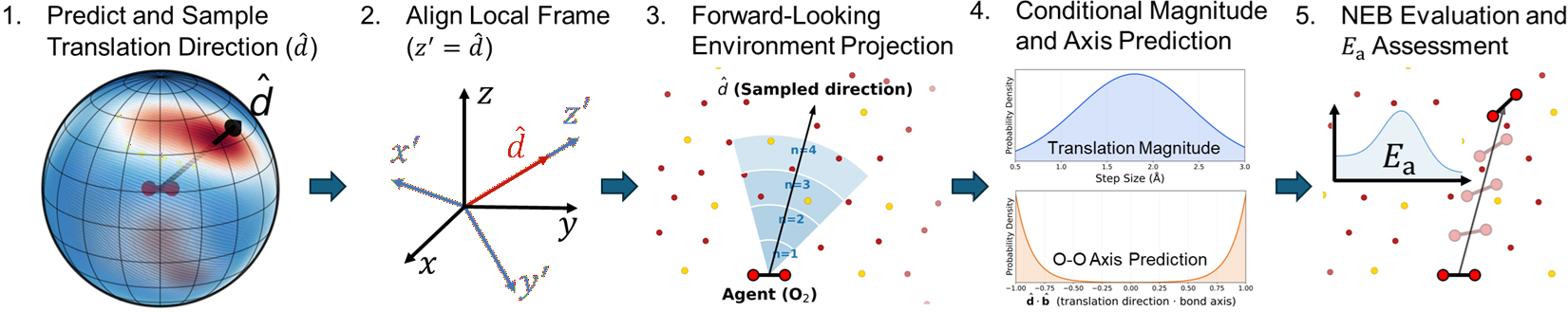}
    \caption{
Autoregressive action generation in the policy network. The policy first samples a translation direction $\hat{\mathbf{d}}$ on $S^2$. The equivariant features are then rotated into a local frame whose $z$-axis is aligned with the sampled direction. From this aligned representation, a forward-looking radial embedding is constructed and used together with aligned scalar features to predict the translation magnitude. In parallel, the O--O axis head predicts the directed bond orientation conditioned on the same sampled direction. This autoregressive design factorizes the action as $p(\hat{\mathbf{d}}, s, \hat{\mathbf{b}})=p(\hat{\mathbf{d}})\,p(s\mid\hat{\mathbf{d}})\,p(\hat{\mathbf{b}}\mid\hat{\mathbf{d}})$ and reflects the physical coupling between where the molecule moves, how far it moves, and how it is oriented during that move.
}
    \label{fig:auto}
\end{figure}

The policy network predicts a continuous rigid-body action for the O$_2$ molecule, consisting of a translation direction $\hat{\mathbf{d}}$, a translation magnitude $s$, and a bond-axis orientation $\hat{\mathbf{b}}$. Rather than predicting these components independently, we use an autoregressive factorization:
\begin{equation}
p(\hat{\mathbf{d}}, s, \hat{\mathbf{b}})
=
p(\hat{\mathbf{d}})
\, p(s \mid \hat{\mathbf{d}})
\, p(\hat{\mathbf{b}} \mid \hat{\mathbf{d}}),
\end{equation}
so that the translation-magnitude and bond-axis heads are conditioned on the sampled translation direction. This reflects the physical coupling between where the molecule moves, how far it moves, and how it is oriented during that move.

This factorization is motivated by the structure of the problem. The translation direction is the primary coarse decision because it determines which local pathway the agent attempts to explore. Once $\hat{\mathbf{d}}$ is chosen, the physically appropriate step size depends on the environment along that direction, such as whether the molecule faces open free volume, a constricted channel, or a nearby reactive site. The preferred O--O axis orientation also depends on the chosen translation direction because the molecule may need to align differently for insertion, approach, or dissociation along different local pathways. Conditioning the downstream predictions on $\hat{\mathbf{d}}$ therefore provides a more physically meaningful parameterization than predicting all action components independently.

Operationally, the sampled direction defines a local frame used to condition the downstream heads. After sampling $\hat{\mathbf{d}}$, the equivariant features are rotated into a frame whose $z$-axis is aligned with $\hat{\mathbf{d}}$. The translation-magnitude head and O--O axis head then make their predictions from features expressed in this direction-aligned frame. In this way, the direction head first selects a candidate pathway, and the subsequent heads refine how that pathway is traversed.

\subsubsection{Translation direction}
The translation-direction head maps the actor state to a probability distribution on the unit sphere $S^2$. The actor state has irreducible-representation structure $128\times 0e + 64\times 1e + 32\times 2e$. Before constructing the spherical distribution, this high-dimensional equivariant state is compressed through learned equivariant projections with gated activations into a lower-dimensional latent representation,
\begin{equation}
16\times 0e + 8\times 1e + 4\times 2e.
\end{equation}
This compression preserves the angular-order structure while reducing the number of multiplicity channels. Because equivariant linear maps preserve angular order, a tensor product is then used to generate higher-order angular features.

A cross-multiplicity tensor-product module then splits the multiplicity copies of each angular order into two halves and computes their products. The highest-order coupling is $2 \otimes 2 \to 4$, giving $\ell_{\max}=4$. Because differently oriented directional features are combined, the resulting spherical-harmonic expansion naturally contains non-zonal ($m \neq 0$) components, allowing the distribution to represent localized off-axis insertion channels rather than only axisymmetric angular patterns. Each tensor-product output is projected to a unified irreducible-representation space and added to a direct equivariant linear path.

A final linear projection produces $(\ell_{\max}+1)^2 = 25$ spherical-harmonic coefficients $\{c_{\ell m}\}$. Because the isotropic term $c_{00}Y_{00}$ is constant on $S^2$, it cancels exactly in the normalized log-density and carries no directional information. We therefore work with the purely directional energy field
\begin{equation}
u(\hat{\mathbf{n}})
=
\sum_{\ell=1}^{\ell_{\max}} \sum_{m=-\ell}^{\ell}
\tilde{c}_{\ell m}\, Y_{\ell m}(\hat{\mathbf{n}}),
\qquad \hat{\mathbf{n}} \in S^2,
\end{equation}
where the $\ell=0$ term has been removed and $\tilde{c}_{\ell m}$ denotes the scaled coefficients. In practice, the raw coefficients are evaluated on a quadrature grid, the weighted mean is subtracted, and the resulting zero-mean field is RMS-normalized so that a learnable inverse temperature
\begin{equation}
\beta = \mathrm{clamp}\!\left(\exp(\log \beta), \beta_{\min}, \beta_{\max}\right)
\end{equation}
controls the sharpness of the directional distribution independently of the raw output scale. The $\ell=0$ coefficient of the scaled coefficient vector is then set to zero, ensuring consistency between the continuous energy field used for Langevin sampling and the grid-evaluated field used for partition-function estimation.

\subsubsection{Langevin sampling on $S^2$}
Given the directional energy field, we sample $\hat{\mathbf{d}}$ using score-guided Langevin dynamics constrained to the sphere rather than selecting a discrete grid point.~\cite{girolami2011riemann} A categorical draw from the grid distribution provides a warm-start initialization, after which the sample is refined by tangent-plane-projected score updates:
\begin{equation}
\hat{\mathbf{n}}_{k+1}
=
\mathrm{normalize}\!\left(
\hat{\mathbf{n}}_k
+
\eta\, \Pi_{T_{\hat{\mathbf{n}}_k}S^2}\nabla_{\hat{\mathbf{n}}} u(\hat{\mathbf{n}}_k)
+
\Pi_{T_{\hat{\mathbf{n}}_k}S^2}\boldsymbol{\xi}_k
\right),
\end{equation}
where $\eta$ is the step size, $\Pi_{T_{\hat{\mathbf{n}}}S^2}$ denotes projection onto the tangent plane of $S^2$ at $\hat{\mathbf{n}}$, and $\boldsymbol{\xi}_k$ is isotropic Gaussian noise projected onto that tangent plane. The score $\nabla_{\hat{\mathbf{n}}}u$ is computed by automatic differentiation of the spherical-harmonic energy field at the current point. This yields a continuous sampled direction $\hat{\mathbf{d}}$ that is not locked to the grid.

The corresponding log-probability is evaluated at the exact sampled direction,
\begin{equation}
\log p(\hat{\mathbf{d}})
=
u(\hat{\mathbf{d}}) - \log Z,
\end{equation}
where $\log Z$ is obtained by quadrature over the grid. During PPO updates, the energy field is recomputed with current network parameters and evaluated at the stored direction, yielding consistent policy ratios without storing grid indices. This strategy was important in practice because, early in training, when the learned angular field is nearly flat, score-guided refinement concentrates samples in higher-energy regions and produces more informative PPO gradients for the directional head.

\subsubsection{Frame alignment}
After sampling $\hat{\mathbf{d}}$, we rotate the compressed equivariant features into a local frame whose $z$-axis is aligned with $\hat{\mathbf{d}}$. Let $R_{\mathrm{align}} \in SO(3)$ denote the rotation from the global frame to this direction-aligned frame. The irreducible components of the feature tensor are transformed with the corresponding Wigner-$D$ matrices:
\begin{equation}
\mathbf{h}^{(\ell)}_{\mathrm{aligned}}
=
D^{(\ell)}(R_{\mathrm{align}}^{\mathsf T})\, \mathbf{h}^{(\ell)},
\end{equation}
so that scalar channels remain unchanged while $\ell>0$ channels rotate consistently.

From the aligned representation we extract only scalar quantities for the downstream invariant heads. Specifically, we retain the 16 original $\ell=0$ scalars, the $z'$-components of the 8 rotated $\ell=1$ vectors, and the $m=0$ components of the 4 rotated $\ell=2$ tensors, for a total of $16+8+4=28$ directional scalars. These quantities are invariant once the frame has been fixed by $\hat{\mathbf{d}}$, so the autoregressive dependence on the sampled direction is introduced through the frame itself rather than by concatenating raw direction vectors.

\subsubsection{Translation magnitude}
The translation magnitude $s$ is predicted conditionally on the sampled direction using only scalar features in the aligned frame. Because $s$ is rotation-invariant once $\hat{\mathbf{d}}$ has been chosen, its predictor operates on invariant inputs rather than the full equivariant representation. These inputs include the directional scalars extracted from the aligned frame, together with a forward-looking radial embedding that provides the magnitude head with information about the environment ahead along the sampled direction.

To construct this radial embedding, we consider each substrate atom $j$ within a cutoff radius $r_{\mathrm{cut}}$ and evaluate a set of spherical Bessel basis functions modulated by a cosine cutoff envelope. Each contribution is additionally weighted by $[\max(0,\, \hat{\mathbf{r}}_j \cdot \hat{\mathbf{d}})]^4$, where $\hat{\mathbf{r}}_j$ is the unit vector from the O$_2$ center of mass to atom $j$. This fourth-power cosine weighting concentrates sensitivity into a narrow forward cone along $\hat{\mathbf{d}}$, while strongly suppressing contributions from atoms located behind or lateral to the proposed direction of motion. The resulting 32-dimensional radial fingerprint is concatenated with the aligned-frame directional scalars before entering the rational-quadratic spline(RQS) conditioner network. In this way, the magnitude head can distinguish whether the sampled direction points into open free volume, toward a nearby reactive site, or into a constricted channel.

The step size is sampled from a one-dimensional normalizing flow based on a  RQS transformation,~\cite{durkan2019neural} which maps a truncated Gaussian base distribution to the physical interval
\begin{equation}
s \in [s_{\min}, s_{\max}].
\end{equation}
A conditioner multilayer perceptron predicts the spline parameters from the scalar inputs, and the change-of-variables formula gives the exact conditional log-probability:
\begin{equation}
\log p(s \mid \hat{\mathbf{d}})
=
\log p_0(z)
+
\log \left| \frac{dz}{ds} \right|,
\end{equation}
where $z$ is the corresponding point in the base space and $p_0$ is the base density. This parameterization yields a smooth bounded distribution with exact log-probabilities, allowing the magnitude head to learn state-dependent step-size preferences while remaining consistent with the autoregressive factorization.

\subsubsection{O--O axis orientation}
Because O$_2$ has $D_{\infty h}$ symmetry, its orientation is fully specified by its bond axis. We therefore predict a directed unit vector $\hat{\mathbf{b}}$ rather than a full rotation matrix. The directionality is important because NEB interpolation preserves atom labels, so $\hat{\mathbf{b}}$ and $-\hat{\mathbf{b}}$ can lead to different interpolated pathways and different apparent barriers. To encode this label-sensitive information, the axis head uses ASYM rather than SYM. The axis prediction is also conditioned autoregressively on the sampled translation direction through the aligned frame.

The aligned-frame features and the full equivariant ASYM embedding are projected into a common equivariant latent space and processed by an equivariant network that outputs two quantities: an equivariant correction vector and an invariant concentration parameter for a von Mises--Fisher (vMF) distribution on the sphere.~\cite{fisher1953dispersion} To obtain a physically meaningful initial mean direction, we construct a geometric prior from the current O--O bond direction and the sampled translation direction in the aligned frame. The learned equivariant correction is projected onto the tangent plane of this prior and mapped to the sphere through the exponential map, yielding a mean direction $\boldsymbol{\mu}_{\hat{\mathbf{b}}} \in S^2$. The final conditional distribution is
\begin{equation}
p(\hat{\mathbf{b}} \mid \hat{\mathbf{d}})
=
C_3(\kappa)\, \exp\!\left(\kappa\, \boldsymbol{\mu}_{\hat{\mathbf{b}}}^{\mathsf T}\hat{\mathbf{b}}\right),
\end{equation}
where
\begin{equation}
C_3(\kappa) = \frac{\kappa}{4\pi \sinh \kappa}
\end{equation}
is the normalizing constant of the three-dimensional vMF distribution and $\kappa$ is the predicted concentration. This construction gives a smooth, directed, symmetry-consistent axis predictor that is initialized near geometrically reasonable orientations but can deviate from them as training discovers lower-barrier pathways.

\subsection{Value network (critic)}
The critic predicts a scalar value from rotation-invariant features only. It does not generate actions and therefore does not need to preserve directional equivariance in its output. In our implementation, the critic input consists of two parts: (i) the $\ell=0$ scalar block of SYM from the frozen SevenNet backbone, and (ii) a small set of hand-constructed geometric invariants describing the current O$_2$ configuration relative to the local environment. These geometric invariants include the O--O bond length, distances from the O atoms and O$_2$ center of mass to nearby Si atoms, a nearest-neighbor distance to the environment, alignment measures between the current bond axis and the nearest-Si direction, a normalized step index within the episode, and invariant norms of the environmental multipole components. Because all of these inputs are scalars, the critic output is strictly invariant under global rigid motions.

The critic itself is a multilayer perceptron that maps the concatenated invariant feature vector to a scalar value estimate,
\begin{equation}
V(\mathbf{X}_t) \in \mathbb{R}.
\end{equation}
The backbone remains frozen, and gradients update only the critic parameters. This separation is useful in practice because the actor requires a richer equivariant representation for directional control, whereas the critic only needs a stable invariant summary that correlates with long-horizon pathway quality.

\subsection{Training algorithm: PPO}
We optimize the policy and critic with Proximal Policy Optimization (PPO).~\cite{schulman2017proximal,schulman2015high} For each accepted step, the rollout buffer stores the state, sampled action components $(\hat{\mathbf{d}}_t, s_t, \hat{\mathbf{b}}_t)$, the corresponding joint log-probability, the value estimate, and the auxiliary information needed to reconstruct the policy during the update, including the environmental multipole used by the actor. The total action log-probability factorizes as
\begin{equation}
\log \pi_{\theta}(\mathbf{a}_t \mid \mathbf{X}_t)
=
\log p_{\theta}(\hat{\mathbf{d}}_t \mid \mathbf{X}_t)
+
\log p_{\theta}(s_t \mid \hat{\mathbf{d}}_t, \mathbf{X}_t)
+
\log p_{\theta}(\hat{\mathbf{b}}_t \mid \hat{\mathbf{d}}_t, \mathbf{X}_t).
\end{equation}
Returns and advantages are computed with generalized advantage estimation.

Given an old policy $\pi_{\theta_{\mathrm{old}}}$ and current policy $\pi_{\theta}$, PPO uses the clipped surrogate objective
\begin{equation}
L_{\mathrm{clip}}(\theta)
=
\mathbb{E}_t\left[
\min\!\left(
r_t(\theta)\,\hat{A}_t,\;
\mathrm{clip}\!\left(r_t(\theta),1-\epsilon,1+\epsilon\right)\hat{A}_t
\right)
\right],
\end{equation}
with
\begin{equation}
r_t(\theta)
=
\frac{\pi_{\theta}(\mathbf{a}_t\mid \mathbf{X}_t)}
{\pi_{\theta_{\mathrm{old}}}(\mathbf{a}_t\mid \mathbf{X}_t)}
=
\exp\!\left(
\log \pi_{\theta}(\mathbf{a}_t\mid \mathbf{X}_t)
-
\log \pi_{\theta_{\mathrm{old}}}(\mathbf{a}_t\mid \mathbf{X}_t)
\right).
\end{equation}
The full training objective combines this clipped policy term with a value loss and an entropy bonus. In practice, we additionally use separate gradient clipping for the directional spherical-harmonic parameters and the remaining network blocks, because the Langevin-based directional sampler can otherwise dominate the optimization dynamics. We also normalize advantages before the PPO update.

\subsection{Reward design}
Reward design balances the primary chemical objective, verified O$_2$ dissociation, against kinetic plausibility quantified through the barriers encountered along the accepted pathway. We therefore use an episode-level reward rather than a purely local step reward:
\begin{equation}
R_{\mathrm{tot}} = w_{\mathrm{diss}} R_{\mathrm{binary}} + w_{\mathrm{Ea}} R_{\mathrm{Ea}}.
\end{equation}
Here $R_{\mathrm{binary}}$ rewards verified dissociation and $R_{\mathrm{Ea}}$ rewards low effective activation barriers. Small step-level shaping or penalty terms can also be included for numerical robustness, such as invalid-step penalties or mild energy-change shaping, but the dominant learning signal is the final episode-level outcome.

\subsubsection{Binary dissociation reward}
The binary term distinguishes successful from unsuccessful episodes after the dissociation-verification protocol described below. If O$_2$ is verified to dissociate, the episode receives a positive terminal reward; if dissociation is not achieved within the finite horizon, the episode receives a penalty. This binary term ensures that the learned policy remains focused on the target chemical event rather than only minimizing barriers for trajectories that never react.

\subsubsection{Effective activation energy}
Let $\{E_i\}_{i=1}^{N}$ denote the set of stepwise activation barriers estimated by NEB for the accepted transitions within an episode, where $E_i$ is the barrier for step $i$ and $N$ is the number of accepted steps. We define an effective activation energy through the log-sum-exp aggregation
\begin{equation}
E_{\mathrm{a,eff}}
=
k_{\mathrm{B}} T
\ln\!\left(
\sum_{i=1}^{N}
\exp\!\left(\frac{E_i}{k_{\mathrm{B}}T}\right)
\right),
\end{equation}
where $k_{\mathrm{B}}$ is the Boltzmann constant and $T$ is the effective temperature parameter. Here we use $T=\SI{1000}{K}$, consistent with typical temperatures used in thermal dry oxidation of silicon.

We convert $E_{\mathrm{a,eff}}$ into a barrier reward using a Fermi--Dirac-type mapping:
\begin{equation}
R_{\mathrm{Ea}}
=
\frac{1}{1 + \exp\!\left(\dfrac{E_{\mathrm{a,eff}}-\mu}{k_{\mathrm{B}}T}\right)}
\, \gamma^{N},
\end{equation}
where $\mu$ is a reference barrier energy and $\gamma \in (0,1]$ is a per-step discount factor that penalizes longer pathways. Since $k_{\mathrm{B}}T \approx \SI{0.086}{eV}$ at \SI{1000}{K}, this form provides a smooth, bounded reward with a physically meaningful thermal scale. The parameter $\mu$ defines the reference point of the reward, where pathways with $E_{\mathrm{a,eff}} < \mu$ receive higher reward and pathways with $E_{\mathrm{a,eff}} > \mu$ receive lower reward.

During training, dissociated O$_2$ molecules remain in the environment, so the environment itself gradually evolves until a threshold number of O$_2$ molecules has dissociated. We refer to the training conducted before resetting the environment as one cycle. The initial value of $\mu$ is set to \SI{1.6}{eV}, which is in the empirically observed range of $E_{\mathrm{a,eff}}$ for the untrained model. After each cycle, $\mu$ is decreased by \SI{0.1}{eV}. In this way, the reward criterion is progressively tightened over training, so that the agent is increasingly encouraged to discover pathways with lower effective activation barriers.

\subsection{Dissociation detection and verification}
\begin{figure}[htbp]
    \centering
    \includegraphics[width=6cm]{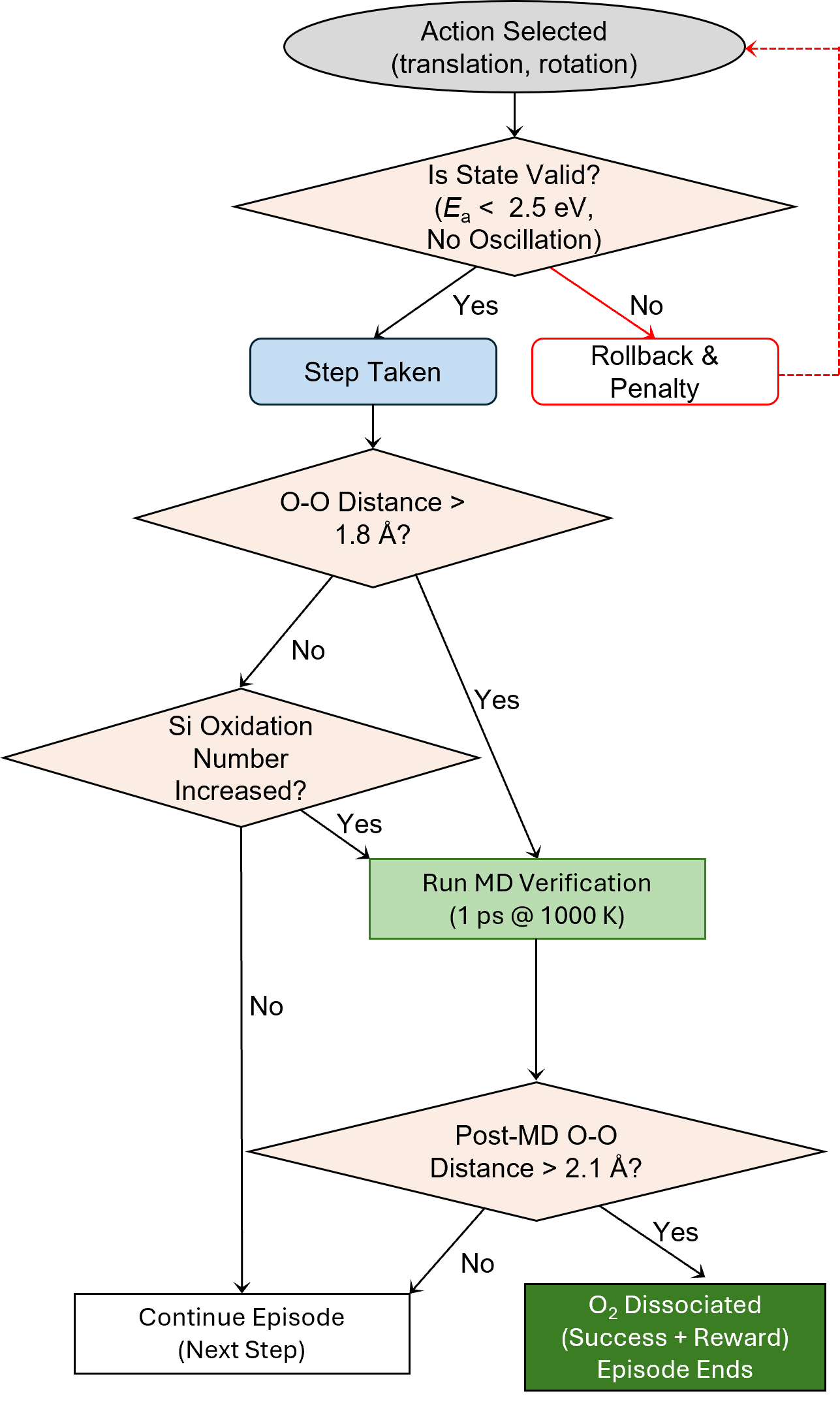}
\caption{
Dissociation detection and verification mechanism. After each accepted step, the environment first checks whether the O--O distance exceeds an initial geometric threshold. If not, an auxiliary oxidation criterion based on the local increase in Si oxidation state near the oxygen atoms is evaluated. A step is flagged as an initial dissociation candidate if either criterion is satisfied. Candidate events are then subjected to short molecular-dynamics verification followed by minimization. Dissociation is accepted only if the O--O distance remains above the stricter final threshold after this verification protocol. This multi-stage procedure rejects transient bond stretching and other false positives, ensuring that only robust oxidation events are counted as successful dissociations.
}
    \label{fig:dissociation}
\end{figure}
A candidate dissociation event is not accepted based on a single geometric criterion alone. Instead, dissociation is identified through a multi-stage procedure designed to reject transient bond stretching and other false positives. After each step, we first test whether the O--O distance exceeds an initial threshold. If it does not, we additionally evaluate an oxidation criterion based on the local increase in Si oxidation state near the oxygen atoms. A step is flagged as an initial dissociation candidate if either the O--O distance criterion or the oxidation criterion is satisfied.

Any such candidate is then subjected to short MD verification followed by minimization. In the current implementation this protocol consists of heating from \SI{300}{K} to \SI{1000}{K}, holding at \SI{1000}{K}, cooling back to \SI{300}{K}, and then relaxing the structure. Dissociation is accepted only if, after this verification protocol, the O--O distance remains above a stricter final threshold of \SI{2.1}{\angstrom}. This conservative final cutoff was chosen because the ReaxFF O--O bond interaction becomes negligible only at somewhat larger separations, so requiring \SI{2.1}{\angstrom} helps ensure that the molecule is genuinely dissociated rather than only transiently elongated.~\cite{noaki2023development} Throughout all stages of the environment update, including rigid-body movement of the O$_2$ molecule, structural relaxation, NEB calculations, and MD verification, the bottom substrate layers are kept frozen to provide a bulk-like region.

\subsection{NEB calculations and rollback}
After each valid accepted step, we run an NEB calculation between the pre-step and post-relaxation local minima to estimate the activation barrier associated with that transition. The resulting barrier contributes to the episode-level set $\{E_i\}$ used in the effective activation energy defined above. In this way, the reward depends not only on whether the final dissociation event occurs, but also on whether the full trajectory reaches that event through kinetically plausible transitions.

If the NEB calculation fails, indicates a numerically unreliable transition, or yields a barrier exceeding a rejection threshold, the environment rolls back to the pre-step configuration and applies a penalty. This rollback mechanism couples pathway discovery to a verification stage, preventing the policy from exploiting simulation artifacts, irreproducible transitions, or kinetically implausible moves. It therefore acts as a physical consistency filter on top of the learned control policy.

\section{Results}
 
\subsection{Training metric}

\begin{figure}[htbp]
  \centering
  \includegraphics[width=\textwidth]{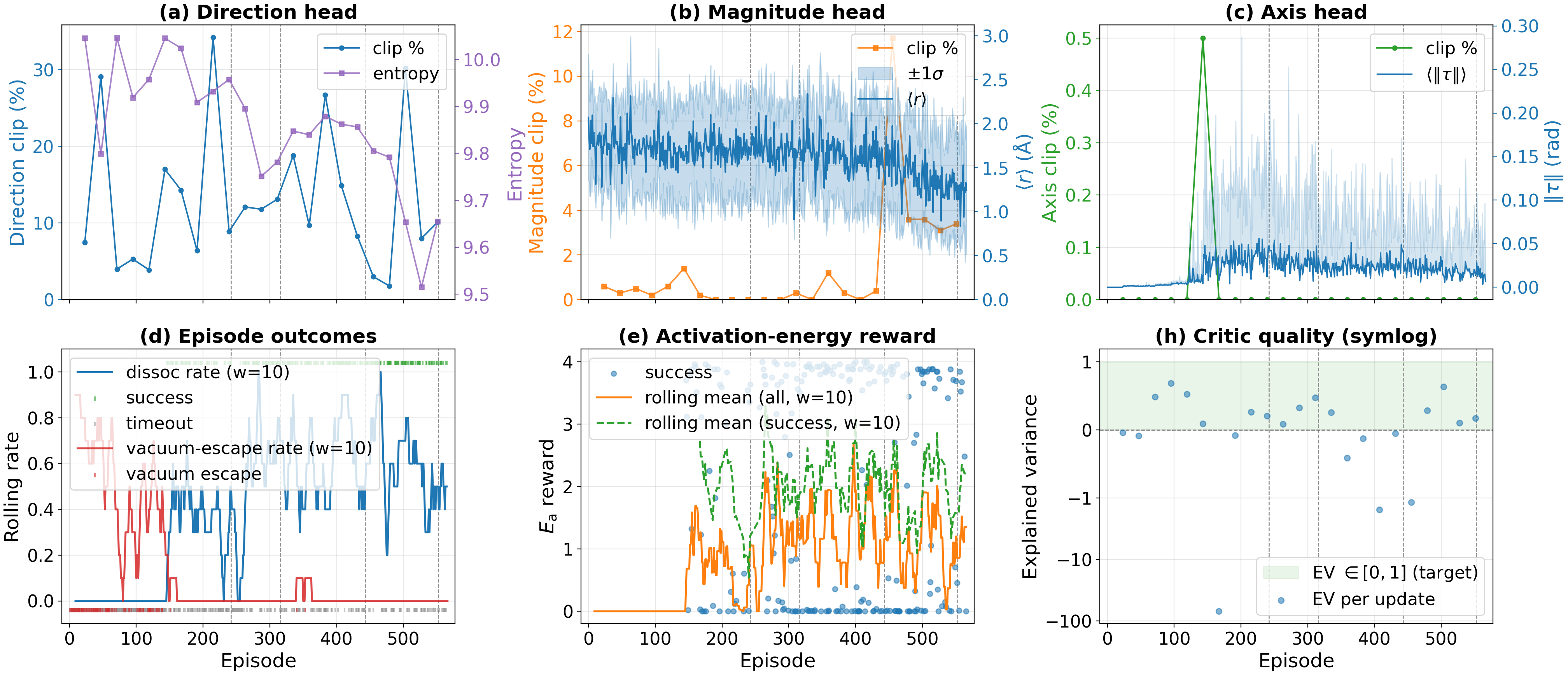}
  \caption{
  Training dynamics. \textbf{(a)--(c)} Per-head policy convergence diagnostics for direction, magnitude, and axis components of the autoregressive action distribution. PPO clip fractions are reported on the left axis of each panel. Per-head behavioural quantities, including entropy, mean step size $\langle r\rangle$ with $\pm 1\sigma$ band, and the $\|\boldsymbol{\tau}\|$ correction from the geometric prior, are reported on the right axis. \textbf{(d)} Episode outcomes, showing rolling rates of MD-verified dissociation (blue) and O$_2$ escape to vacuum (red), with per-episode markers along the top and bottom edges. \textbf{(e)} Activation-energy reward $R_{Ea}$ per episode, shown as blue points for dissociation-verified episodes only, together with rolling means over all episodes (orange) and success-only episodes (green dashed). \textbf{(h)} Critic explained variance per PPO update on a symmetric-log $y$-axis. The shaded green band marks the $\mathrm{EV}\in[0,1]$ target regime. Dashed vertical lines mark the end of each completed cycle.
  }
  \label{fig:training}
\end{figure}

Figure~\ref{fig:training} summarizes the training run on the Si/$a$-SiO$_2$ environment. Figure~\ref{fig:training}~(a)--(c) track convergence of the three policy heads in the autoregressive factorization $p(\hat{\mathbf{d}})\,p(s\mid\hat{\mathbf{d}})\,p(\hat{\mathbf{b}}\mid\hat{\mathbf{d}})$. Figure~\ref{fig:training}~(d) shows episode outcomes, Fig.~\ref{fig:training}~(e) shows the activation-energy reward, and Fig.~\ref{fig:training}~(h) shows critic quality. Figure~\ref{fig:cycle} collects the corresponding per-cycle physical metrics computed from MD-verified dissociation episodes.

We refer to the early part of training as the warm-up phase. At the beginning of training the policy is randomly initialized, so the rigid-body moves proposed for O$_2$ are only weakly related to the local atomistic environment. Under this policy, MD-verified dissociation is rare, and the activation-barrier reward $R_{\mathrm{Ea}}$ is only defined for episodes that contain at least one verified dissociation event. The agent therefore receives no pathway-quality signal until it first reaches dissociation by chance. Before that point, learning is driven almost entirely by the binary terminal outcome, so exploration is effectively unguided.

The strongest convergence signal appears in the direction head, shown in Fig.~\ref{fig:training}a. Its PPO clip fraction is high during the warm-up phase, with early values near 30 \%, and then falls toward the (5 to 10)~\% range later in training. Over the same interval, the entropy decreases by about 0.4 nats. This is the clearest sign that the policy is learning a more selective directional distribution rather than continuing to sample nearly isotropically.

The magnitude head changes more gently, as shown in Fig.~\ref{fig:training}(b). The mean step size drifts downward only slightly over the full run, while the $\pm 1\sigma$ spread remains around $\pm 0.6$~\AA. Its clip fraction stays close to zero for most of training, aside from a single late spike. This suggests that optimization of the step-size distribution is comparatively mild and that most of the policy adaptation occurs through directional choice rather than through large changes in the displacement scale.

The axis head also remains stable throughout training. In Fig.~\ref{fig:training}(c), the correction norm $\|\boldsymbol{\tau}\|$ rises from essentially zero during warm-up to a small quasi-stationary value of about 0.03~rad after dissociation events begin to appear. The learned orientation update is therefore modest, but not negligible. The axis clip fraction remains near zero throughout. Taken together, Fig.~\ref{fig:training}~(a)--(c) show that the policy updates are numerically well behaved and that no single head dominates the PPO clipping dynamics in an unstable way.

The per-episode outcomes in Fig.~\ref{fig:training}(d) show a clear behavioral shift. During roughly the first 150 episodes, the policy frequently drives O$_2$ away from the surface and into vacuum, while no successful dissociations are observed. Once the first verified dissociation appears near episode 145, the vacuum-escape rate collapses toward zero and the rolling dissociation rate rises into the 0.4 to 0.8 range. From that point onward, the policy is no longer learning only how to avoid obvious failure. It is learning in the regime where successful oxidation events occur often enough to shape the pathway distribution.

The activation-energy reward in Fig.~\ref{fig:training}(e) follows the same transition. Rewards remain at zero during the warm-up because no verified dissociation occurs. After dissociation begins, the success-conditioned rolling mean rises rapidly and then stays mostly in the range 1.5--2.5. This indicates that, once the policy begins to reach the interface productively, barrier-sensitive learning becomes active and remains relevant throughout the rest of training.

The critic explained variance is shown in Fig.~\ref{fig:training}(h) on a symlog axis. A single early excursion to a large negative value occurs immediately after the onset of dissociation rewards. This is expected as the critic has first been trained on an almost degenerate zero-return signal and then is suddenly exposed to positive-return episodes. After this transition, the explained variance recovers into the target band and stays mostly between 0 and 1. A second, smaller excursion appears in the late-training region. This coincides with the rise in the magnitude-head clip fraction in panel~(b) and the contraction of $\langle r\rangle$ toward smaller values. We interpret this dip as the critic adapting to a qualitative shift in policy behavior, which is documented and discussed in connection with Sec.~\ref{fig:Per-cycle metrics}.

\subsection{Per-cycle metrics}
\label{fig:Per-cycle metrics}
  
\begin{figure}[htbp]
  \centering
  \includegraphics[width=\textwidth]{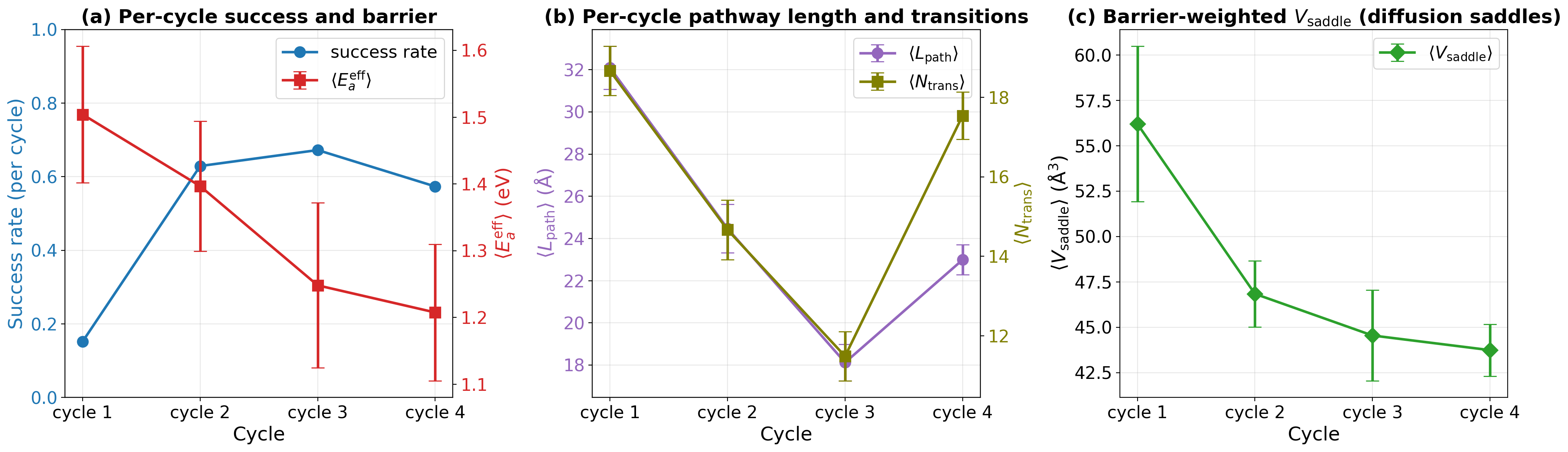}
  \caption{
  \textbf{(a)} Per-cycle success rate (left axis, blue) and mean effective activation barrier $\langle E_a^{\rm eff}\rangle$ (right axis, red). \textbf{(b)} Per-cycle mean pathway length $\langle L_{\rm path}\rangle$ (left, purple) and mean number of accepted NEB transitions $\langle N_{\rm trans}\rangle$ (right, olive). \textbf{(c)} Barrier-importance-weighted mean local void volume at NEB saddle structures, $\langle V_{\rm saddle}\rangle$, for diffusion-step transitions. Error bars show the standard error of the mean (SEM = $\sigma/\sqrt{n}$), where $n$ is the number of MD-verified dissociating episodes contributing to a given cycle. The within-cycle 1$\sigma$ spreads of per-episode quantities are reported separately in the text. Per-step weights are $w_i = \exp(E_i/k_{\mathrm{B}}T) / \sum_j \exp(E_j/k_{\mathrm{B}}T)$ summed over the full barrier set of each episode, so that each step contributes in proportion to its weight in the episode's $E_a^{\rm eff}$.
  }
  \label{fig:cycle}
\end{figure}

The training run produced four complete oxidation cycles. Each cycle ended when the substrate-reset condition was reached, namely incorporation of 80 new O atoms into the Si/$a$-SiO$_2$ environment. The per-cycle metrics in Fig.~\ref{fig:cycle} are computed only over MD-verified dissociation episodes.

\subsubsection{Episode efficiency}

The success rate rises sharply between cycle~1 and cycle~2, from 0.15 to 0.63, increases more modestly to 0.67 in cycle~3, and then declines slightly to 0.57 in cycle~4, as shown in Fig.~\ref{fig:cycle}(a). The first cycle still contains much of the warm-up regime, where the policy is learning basic surface retention and many episodes end without verified dissociation. By cycle~2, the policy has largely overcome that failure mode and the success rate rises sharply. The smaller gain into cycle~3 suggests that the dominant remaining failures are no longer simple exploration failures, but instead reflect the intrinsic difficulty of the local dissociation environment. The slight decline in cycle~4 coincides with a qualitative change in policy behavior, discussed below, in which the policy adopts a finer-grained sampling strategy that does not directly maximize immediate success rate.

\subsubsection{Effective activation barrier}

The cycle-averaged effective activation barrier decreases monotonically across all four cycles, from ($1.50 \pm 0.10$)~eV in cycle~1 to ($1.40 \pm 0.10$)~eV in cycle~2, ($1.25 \pm 0.12$)~eV in cycle~3, and ($1.21 \pm 0.10$)~eV in cycle~4 (mean $\pm$ SEM, $n = 40$ dissociating episodes per cycle; Fig.~\ref{fig:cycle}(a)). The cycle~1 to cycle~3 reduction of about 0.25~eV is substantially larger than the combined SEM across cycles, indicating that the decrease reflects a genuine reduction in the cycle-mean barrier rather than statistical fluctuation. This downward trend is important because it shows that improved success is not achieved simply by finding more dissociation events of arbitrary quality. The policy is finding dissociation pathways with lower effective barriers throughout training, although the improvement clearly diminishes between cycle~3 and cycle~4.

This behavior is consistent with the reward design. The barrier reward favors lower $E_a^{\rm eff}$, and the reference midpoint $\mu$ is tightened from cycle to cycle according to the fixed schedule described in Sec.~2.10.2. Improvement in this metric therefore indicates that the policy is adapting to an increasingly stringent barrier criterion rather than merely exploiting a fixed threshold.

The absolute scale of these barriers is also consistent with experimental estimates for diffusion-limited dry oxidation of silicon. In this regime, the characteristic activation energy for oxygen transport through a-SiO$_2$ is commonly reported to be about 1.2~eV.~\cite{lamkin1992oxygen,massoud1985thermal,bongiorno2004multiscale} The mean values reached in cycles~3 and~4, $\langle E_a^{\rm eff}\rangle = 1.25$~eV and $1.21$~eV, are therefore in good agreement with the expected transport barrier for diffusion-controlled oxidation. This supports the view that the learned policy is discovering kinetically plausible pathways in the diffusion-limited regime, rather than merely producing dissociation events with arbitrary barrier scale.

\subsubsection{Pathway length and transition count}

The mean pathway length and the mean number of accepted NEB transitions show qualitatively different behavior between cycles~1--3 and cycle~4. In cycles~1--3, both quantities decrease monotonically: $\langle L_{\rm path}\rangle$ goes from ($32.1 \pm 1.0$)~\AA{} to ($24.6 \pm 1.2$)~\AA{} to ($18.1 \pm 0.9$)~\AA{}, and $\langle N_{\rm trans}\rangle$ goes from ($18.7 \pm 0.7$) to ($14.8 \pm 0.7$) to ($11.5 \pm 0.6$), as shown in Fig.~\ref{fig:cycle}(b). The two quantities track one another closely over this interval, and Fig.~\ref{fig:training}(b) shows only a mild downward drift in the mean step size $\langle r\rangle$ over the same range. The shorter total path length in cycles~1--3 is therefore driven mainly by fewer accepted transitions, not by a large change in the size of each step. In physical terms, the learned trajectories become more direct, with the policy reaching dissociation through fewer intermediate moves rather than through substantially larger displacements.

In cycle~4, this trend reverses. Both $\langle L_{\rm path}\rangle$ and $\langle N_{\rm trans}\rangle$ increase, to ($23.0 \pm 0.7$)~\AA{} and ($17.5 \pm 0.6$), respectively. However, the path length grows less than the transition count. The per-step displacement $L_{\rm path}/N_{\rm trans}$ therefore decreases from $1.58$~\AA{}/step in cycle~3 to $1.31$~\AA{}/step in cycle~4. This contraction is reflected directly in the magnitude head in Fig.~\ref{fig:training}(b), where the orange clip-fraction trace rises sharply near the end of training and the $\langle r\rangle$ band tightens at smaller values. Cycle~4 therefore corresponds to a deliberate change in policy behavior rather than a regression. Instead of continuing to shorten paths, the policy adopts a finer-grained mode of motion in which more, smaller steps are taken per episode.

\subsubsection{Local saddle geometry}

To examine the local environments associated with the kinetic bottlenecks along policy-discovered pathways, we compute a local void volume at the NEB saddle image for every accepted diffusion-step transition. The dissociation step itself is excluded. We define
\[
V_{\rm saddle} = \frac{4}{3}\pi R^3,
\]
where $R$ is the distance from the O$_2$ center of mass to the nearest network atom in the saddle-image structure. Steps with barriers below $k_{\mathrm{B}}T$ are excluded.

Because $E_a^{\rm eff}$ is dominated by the largest barriers through the log-sum-exp form, we weight each step by its softmax importance within the full barrier set of its episode. This focuses the analysis on the steps that matter most for the episode-level kinetic cost, rather than on the more numerous low-barrier diffusive moves.

The barrier-weighted $\langle V_{\rm saddle}\rangle$ decreases from ($56.2 \pm 4.3$)~\AA$^3$ in cycle~1 to ($46.8 \pm 1.8$)~\AA$^3$ in cycle~2 and ($44.6 \pm 2.5$)~\AA$^3$ in cycle~3, as shown in Fig.~\ref{fig:cycle}(c). It is tempting to expect that a larger local ring or void at the saddle should correspond to a lower O$_2$ diffusion barrier in $a$-SiO$_2$, because a wider opening would reduce repulsion between the diffusing molecule and the surrounding network. However, real disordered oxides are more complicated than any single geometric metric can capture. First-principles results from Bakos \emph{et al.}~\cite{bakos2002reactions} already show that O$_2$ diffusion barriers are not controlled monotonically by one simple descriptor such as nominal ring size, since substantially different barriers can arise even for six-member Si--O rings of comparable diameter. In this context, our observation that $\langle V_{\rm saddle}\rangle$ decreases while $\langle E_a^{\rm eff}\rangle$ also decreases is not merely a counterintuitive anomaly. Rather, it indicates that kinetically favorable pathways in this system are not determined by a single obvious geometric measure. More broadly, this highlights the value of a human-intuition-agnostic, automatically learned framework such as ours, which can discover favorable transition pathways without requiring the relevant structural descriptor to be specified in advance.

Taken together, the per-cycle trends support a coherent picture. As training proceeds, the policy dissociates O$_2$ more reliably, reaches dissociation through shorter paths with fewer accepted transitions, and lowers the effective barrier of the resulting pathways. These improvements are exactly the combination one would want from a barrier-aware control policy in a rare-event atomistic environment.

\subsubsection{Barrier distribution and policy strategy across cycles}

\begin{figure}[htbp]
  \centering
  \includegraphics[width=\textwidth]{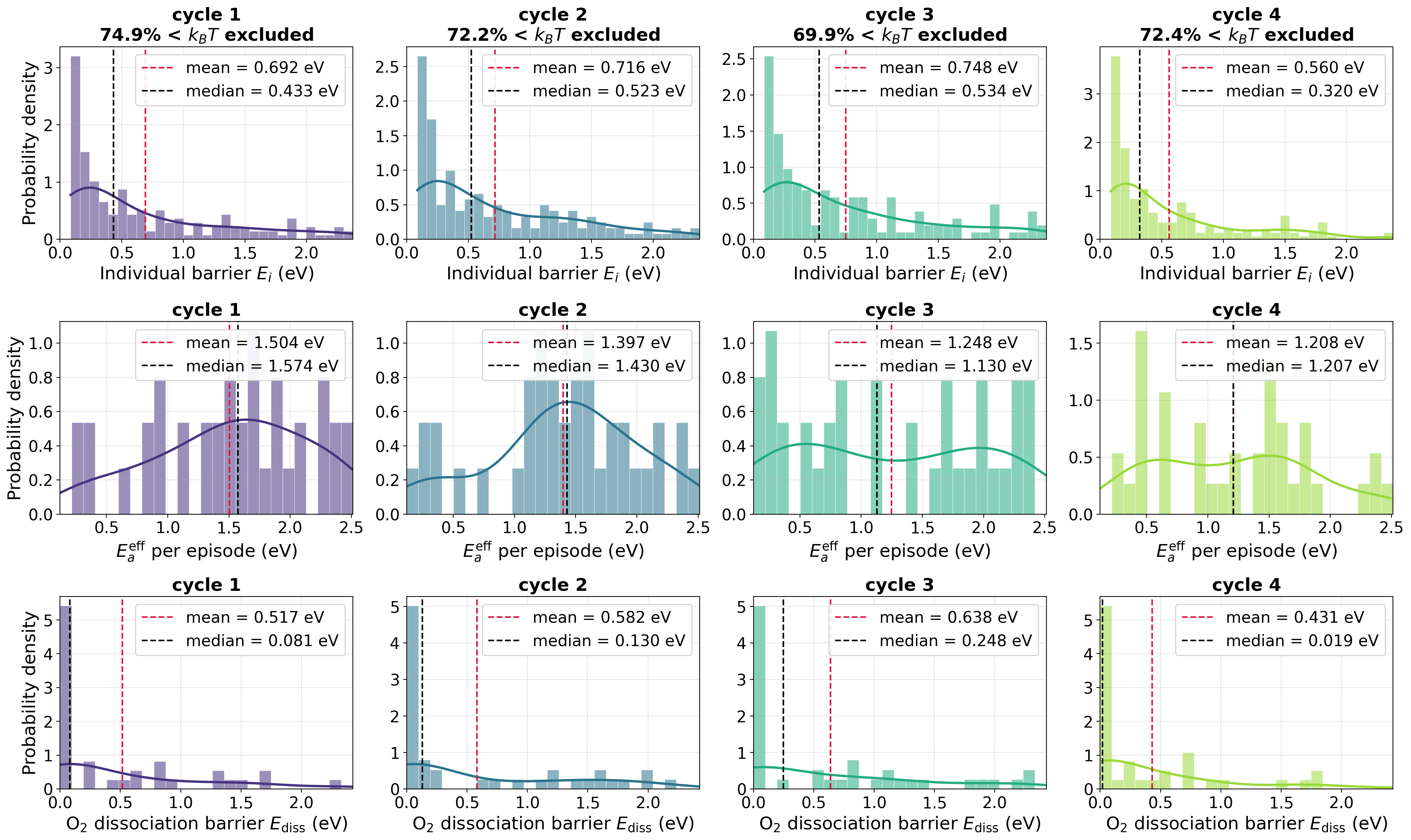}
  \caption{
  Per-cycle distributions of NEB barriers across the four completed oxidation cycles, restricted to MD-verified dissociation episodes. \textbf{(top row)} Distribution of individual NEB barriers $E_i$ along accepted episode trajectories. Only barriers with $E_i > k_{\rm B}T$ at $T=1000$~K ($\approx 0.086$~eV) are shown. The fraction of barriers below this kinetic threshold is annotated above each panel and remains in the (70 to 75)~\% range across cycles. \textbf{(middle row)} Distribution of the effective activation barrier $E_a^{\rm eff}$ per episode. \textbf{(bottom row)} Distribution of the O$_2$ dissociation barrier $E_{\rm diss}$ per episode, defined as the NEB barrier of the final accepted transition in each dissociated episode. This is the barrier of the step whose post-relaxation minimum was MD-verified as dissociated. All preceding barriers in the episode therefore correspond to diffusion steps. Vertical dashed lines mark the mean (red) and median (black). Smooth curves show kernel density estimates of each distribution.
  }
  \label{fig:barrier}
\end{figure}

The three distributions in Fig.~\ref{fig:barrier} capture different aspects of the learned trajectories. The individual-barrier distribution describes the full set of accepted moves within each episode, including both the rate-limiting saddle and the more numerous low-barrier intermediate transitions that connect it. The effective-barrier distribution, by contrast, is dominated by the largest barrier encountered in each path. With $E_i / k_{\rm B}T$ typically in the range of 3--25 for accepted barriers, the soft maximum
\begin{equation}
E_a^{\rm eff} = k_{\rm B}T\,\log \sum_i e^{E_i/k_{\rm B}T} \;\approx\; \max_i E_i + k_{\rm B}T\,\log N_{\rm eff}
\end{equation}
is effectively controlled by $\max_i E_i$, with the logarithmic correction contributing only about 0.2~eV even for the longest paths in our data. Thus, $E_a^{\rm eff}$ is determined primarily by the rate-limiting saddle, whereas the individual-barrier distribution reflects the overall sampling pattern of the policy. The dissociation-barrier distribution isolates the final bond-breaking step, allowing it to be examined separately from the preceding diffusion-step ensemble.

Across cycles~1--3, the pooled distribution of individual step barriers changes only modestly, while $E_a^{\rm eff}$ decreases clearly. This is not contradictory because the two quantities probe different aspects of the trajectories. The pooled barrier distribution summarizes all accepted local moves across episodes, whereas $E_a^{\rm eff}$ is controlled mainly by the largest barrier within each episode. The drop in $E_a^{\rm eff}$ therefore suggests that the policy is increasingly selecting pathways with lower rate-limiting barriers, even though the overall set of locally available barriers in the disordered oxide remains broadly similar.

In cycle~4, the individual-barrier distribution changes more visibly. The median drops from $0.53$~eV in cycle~3 to $0.32$~eV in cycle~4, and the distribution gains substantial weight in the $0.1$ to $0.3$~eV range. This shift is consistent with the strategy change described in the previous subsection. Smaller translation magnitudes and a larger number of accepted transitions per episode generate many low-cost intermediate moves. Because of the soft-maximum behavior above, these moves contribute negligibly to $E_a^{\rm eff}$, which is why the cycle~4 $E_a^{\rm eff}$ distribution remains qualitatively similar to that of cycle~3 even though the underlying motion is different.

As discussed above, the mean effective barrier per cycle decreases from 1.50~eV in cycle~1 to 1.40~eV in cycle~2, 1.25~eV in cycle~3, and 1.21~eV in cycle~4. The per-episode standard deviation of $E_a^{\rm eff}$ remains substantial in all four cycles, with $\sigma \approx 0.6$ to $0.8$~eV. This is physically reasonable because the policy operates in a disordered oxide environment, where the local kinetic landscape varies strongly from episode to episode. The somewhat larger spread in cycle~3, $\sigma = 0.77$~eV, suggests that the policy is sampling a broader set of accessible low-barrier pathways during this phase, whereas the slightly smaller spread in cycle~4, $\sigma = 0.64$~eV, is consistent with the more refined sampling regime.

The dissociation-barrier distribution provides the clearest evidence of what the policy is doing during the cycle~4 strategy shift. The cycle~4 median $E_{\rm diss}$ collapses to $0.019$~eV, the lowest value across all cycles, with a strong concentration of weight near zero. This places a nontrivial constraint on the underlying policy behavior. If the cycle~4 strategy merely added cheap intermediate moves while leaving the dissociation step itself unchanged, then the $E_{\rm diss}$ distribution would remain similar to that of cycle~3, where the median was $0.25$~eV. Instead, the dissociation step itself is being driven toward a nearly barrierless regime. Mechanistically, the smaller and more numerous moves appear to position O$_2$ more precisely at configurations where bond breaking becomes essentially downhill. Once such a configuration is reached, the dissociation step proceeds with negligible kinetic cost, while the rate-limiting barrier of the episode remains in the preceding diffusion segment.

Taken together, the four cycles suggest two distinct learning regimes. In cycles~1--3, the policy primarily improves efficiency by discovering shorter dissociation pathways with fewer accepted transitions and lower rate-limiting barriers, while the pooled distribution of intermediate moves remains broadly similar. In cycle~4, the policy shifts toward a refinement regime in which it uses smaller, more numerous moves to navigate the local environment near the saddle and place O$_2$ into near-barrierless dissociation configurations. This finer-grained motion does not produce a comparably large additional drop in $E_a^{\rm eff}$, because the rate-limiting diffusion barrier remains similar to that of cycle~3. However, the collapse of $E_{\rm diss}$ shows that the underlying microscopic behavior has changed qualitatively. That this transition emerges spontaneously, without any explicit prompting in the reward function or training schedule, indicates that the framework can express distinct solution strategies as the local kinetic landscape evolves across cycles.

\subsection{Post-training evaluation}
\begin{figure}[htbp]
  \centering
  \includegraphics[width=\textwidth]{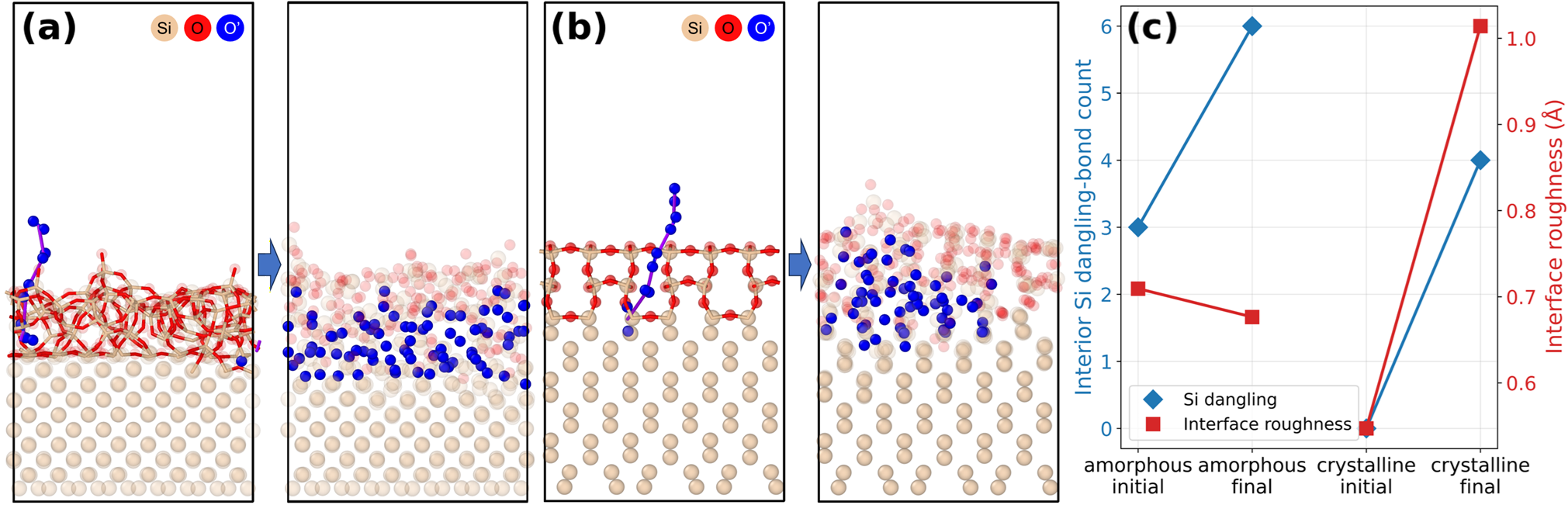}
  \caption{
  Evaluation-mode oxidation trajectories generated by the trained policy without PPO updates. \textbf{(a)} Si/$a$-SiO$_2$ case. Left, initial structure used for evaluation, which is identical to the starting structure used for training. Right, final structure after incorporation of 80 new O atoms. \textbf{(b)} Si/$\beta$-tridymite SiO$_2$ case. Left, initial out-of-training-domain structure used to test transferability. Right, final structure after incorporation of 60 new O atoms. In panels (a) and (b), red spheres denote oxygen atoms initially present in the oxide, blue spheres denote oxygen atoms incorporated during evaluation, and the purple line traces the center-of-mass trajectory of the first dissociating O$_2$ molecule. \textbf{(c)} Interior Si dangling-bond count (left axis, blue diamonds) and interface roughness $R_q$ (right axis, red squares) for the four states shown in panels (a) and (b). A Si atom is counted as a dangling bond if its total coordination number is below four and it has at least one oxygen neighbor.
  }
  \label{fig:Evaluation}
\end{figure}

Once a representative trained policy has been obtained, we perform post-training evaluation with PPO updates disabled. In this mode, the policy is used only for inference, and at each step the action with the highest policy log-probability is selected. This allows the learned policy to be used as a fast pathway generator for oxidation simulations after training. When the goal is primarily to obtain the final oxidized structure at low computational cost, NEB calculations can in principle be omitted once sufficient confidence has been established that the selected actions remain kinetically reasonable. If detailed kinetic information is still desired, NEB calculations can be retained during evaluation. In the present demonstration, we keep NEB enabled so that barrier information can be reported together with the generated oxidation trajectories.

We consider two types of starting structures in post-training evaluation. The first is the Si/$a$-SiO$_2$ structure used during training, shown in Fig.~\ref{fig:Evaluation}(a). The second is a Si/$\beta$-tridymite SiO$_2$ interface, shown in Fig.~\ref{fig:Evaluation}(b). The crystalline case is not intended as a quantitatively realistic model of thermally grown gate oxide. Rather, it is included for three specific reasons. First, because it was not used during training, it provides an out-of-training-domain test of whether the learned policy still behaves in a physically reasonable way. Second, the crystalline geometry makes the early O$_2$ pathway easier to inspect visually than in the amorphous case, where the local free-volume network is much harder to follow by eye. Third, it highlights a broader methodological point. Conventional Si/crystalline-SiO$_2$ models, although convenient and widely used in computational studies of Si/SiO$_2$ interfaces, can lead to oxidation behavior and interface evolution that differ qualitatively from the amorphous case that is more relevant to dry thermal oxidation.

The left panel of Fig.~\ref{fig:Evaluation}(a) shows the initial Si/$a$-SiO$_2$ snapshot used for evaluation. This is the same starting configuration used in the training runs discussed above. The right panel shows the final structure after incorporation of 80 new oxygen atoms. The average effective activation barrier over this evaluation trajectory is \SI{1.16}{eV}. The added oxygen atoms are distributed relatively uniformly in the lateral directions and dissociate predominantly near the Si/$a$-SiO$_2$ interface. This behavior is consistent with the established picture of dry silicon oxidation, in which intact O$_2$ molecules diffuse through the oxide and react near the buried interface.~\cite{rosencher197918o,rochet198418o}

The left panel of Fig.~\ref{fig:Evaluation}(b) shows the initial Si/$\beta$-tridymite SiO$_2$ structure together with the trajectory of the first dissociating O$_2$ molecule. Although this structure lies outside the training set, the learned policy still identifies a physically sensible entry pathway. The molecule first enters an open ring at the crystalline SiO$_2$ surface and reaches a large interstitial void within the $\beta$-tridymite network, which acts as a local minimum site. Relative to the state in which O$_2$ remains in vacuum, this intermediate lowers the energy by \SI{3.28}{eV}. Subsequent motion toward incorporation at the buried Si/c-SiO$_2$ interface lowers the energy further, by \SI{6.51}{eV} relative to the intermediate site. Because this pathway is visually much clearer than in the amorphous case, it provides a useful qualitative check that the learned policy is not moving O$_2$ arbitrarily, but is instead identifying physically sensible entry points and intermediate trapping sites even in an out-of-training-domain structure.

The right panel of Fig.~\ref{fig:Evaluation}(b) shows the final structure after incorporation of 60 new oxygen atoms. In contrast to the Si/$a$-SiO$_2$ case, the newly incorporated oxygen atoms do not remain laterally distributed, but instead accumulate in a more localized region. This suggests that once oxidation begins to perturb the crystalline interface, the resulting locally disordered region provides additional diffusion channels and more favorable dissociation sites, so that subsequent oxidation becomes spatially concentrated around the initially oxidized region. In this sense, the first oxidized region acts as a nucleation point for further growth. The average effective activation barrier in this case is \SI{1.47}{eV}, which is higher than in the Si/$a$-SiO$_2$ evaluation. Together with the more localized oxygen accumulation, this indicates that a crystalline starting oxide can bias both the inferred oxidation pathway and the resulting interface structure.

This interpretation is also consistent with experiment. Buried-interface propagation during silicon oxidation has been directly observed experimentally, indicating that oxidation proceeds through controlled advance of the Si/SiO$_2$ interface rather than arbitrary disordering throughout the oxide.~\cite{ross1992dynamic} In addition, Lai and Irene showed that thermal oxidation drives initially rough and initially smooth Si surfaces toward a limiting Si/SiO$_2$ interface roughness of about \SI{0.3}{nm} rms after extensive oxidation.~\cite{lai1999limiting} Together, these observations suggest that dry thermal oxidation tends to preserve a relatively abrupt and controlled interface morphology even as the buried interface advances.

Motivated by this experimental picture, we next examine two structural quantities closely tied to Si/SiO$_2$ interface quality, namely defect density and interface roughness. As a simple defect metric, we count Si dangling bonds, which are closely related to under-coordinated Si environments associated with electrically active interface traps such as $P_b$ centers in MOS devices.~\cite{futako2004situ,keunen2011inherent} To quantify interface roughness, we follow Cvitkovich \emph{et al.} and partition the $xy$ plane into a $12 \times 12$ lateral grid.~\cite{cvitkovich2023dynamic} In each populated bin, the lowest-$z$ oxygen atom is taken as the local interface position $z_{\rm bin}$. The root-mean-square roughness is then defined as
\begin{equation}
    R_q = \sqrt{\left\langle \left( z_{\rm bin} - \bar{z} \right)^2 \right\rangle},
\end{equation}
where $\bar{z}$ is the mean of $z_{\rm bin}$ over the populated bins.

Figure~\ref{fig:Evaluation}(c) shows that for the Si/$a$-SiO$_2$ starting structure, the dangling-bond count increases during further oxidation while the interface roughness remains nearly unchanged. The starting Si/$a$-SiO$_2$ structure was generated by deposition MD until the oxide thickness converged, corresponding to the early fast-oxidation stage. At that stage, the interface is already oxygen-rich and relatively abrupt. Continued oxidation in the diffusion-controlled regime, sampled here by the REALIZE framework, introduces additional interfacial defects without destroying that overall abrupt interface morphology. In other words, the framework accesses defect-forming oxidation pathways that were not sampled during the initial MD preparation, while still preserving the experimentally expected abrupt-interface character.

This contrasts sharply with the crystalline-starting case, which evolves toward a substantially rougher interface. The comparison therefore suggests that simple thin-oxide models prepared only by MD, as well as idealized Si/crystalline-SiO$_2$ interface models, can misrepresent how continued oxidation modifies interface quality. We note that our atomistically defined $R_q$ is not identical to that experimental roughness measure, so the comparison should be interpreted at the level of overall scale rather than as a one-to-one validation. Even so, it is useful in showing that continued oxidation should be judged against experimentally constrained interface morphology, not only against simplified starting structures.

In our evaluated structures, the corresponding dangling-bond density falls in the range of $10^{13}$--$10^{14}$~cm$^{-2}$, which is substantially higher than the $P_b$-type defect densities typically reported for unpassivated thermal Si/SiO$_2$ interfaces, generally in the $10^{12}$~cm$^{-2}$ range.~\cite{futako2004situ,keunen2011inherent} Although this difference is larger than the experimental values, several factors likely contribute to it. Here we count all under-coordinated Si atoms with at least one oxygen neighbor, rather than only the specific defect configurations identified experimentally as $P_b$ centers. The evaluated structures are also unpassivated, so the measured dangling-bond density should be interpreted as a broader structural defect metric rather than a direct one-to-one estimate of the experimentally reported $P_b$ density. In addition, the use of ReaxFF may contribute quantitative error in the defect statistics. More accurate electronic-structure methods or improved force fields may bring this estimate closer to experimentally reported values.

\section{Discussion}

\subsection{Lower-Bound Timescale Estimate for the Discovered Pathways}
The present framework is designed for goal-directed discovery of kinetically plausible pathways in complex, heterogeneous environments. The learned policy is not intended to reproduce unbiased dynamics. Instead, it optimizes an episode-level objective that rewards verified dissociation together with low effective activation barriers estimated from NEB. Accordingly, RL trajectories should not be interpreted as direct realizations of experimental time evolution. In a real oxidizing environment, an O$_2$ molecule undergoes thermally activated diffusion through a disordered network, repeatedly sampling many micro-pathways before reacting. By contrast, the RL agent is encouraged to reach dissociation within a finite horizon, so prolonged diffusive wandering is suppressed. The discovered trajectories are therefore best viewed as mechanistically meaningful low-barrier pathways and kinetic bottlenecks, together with a realistic evolved interfacial structure at the end of the episode.

Even so, the discovered pathways can still be used to estimate a lower-bound execution time once a productive route has been identified. For a sequence of barriers $\{E_i\}$, transition-state theory gives the stepwise rate
\begin{equation}
k_i \approx \nu_i \exp\left(-\frac{E_i}{k_{\mathrm{B}} T}\right),
\end{equation}
and the corresponding pathway time may be estimated as
\begin{equation}
\tau_{\mathrm{path}} \sim \sum_i k_i^{-1}.
\end{equation}
Here, $\nu_i$ is an event-dependent attempt frequency. For O$_2$ dissociation, the relevant coordinate is O--O stretching and bond breaking, so we estimate the attempt frequency from the intrinsic O--O vibrational frequency. The gas-phase harmonic vibrational frequency of O$_2$ is approximately $1580$~cm$^{-1}$, corresponding to $4.7 \times 10^{13}$~s$^{-1}$.~\cite{nist_webbook_diatomic} We therefore use $\nu_{\mathrm{diss}} = 5 \times 10^{13}$~s$^{-1}$. For O$_2$ diffusion or hopping events, the relevant coordinate is instead associated with frustrated translation, local reorientation, and migration through the disordered oxide network, for which we use a conventional activated-hopping prefactor of $\nu_{\mathrm{diff}} = 1 \times 10^{13}$~s$^{-1}$. Thus,
\begin{equation}
k_i =
\begin{cases}
5 \times 10^{13}
\exp\left(-\frac{E_i}{k_{\mathrm{B}} T}\right), & \text{O$_2$ dissociation}, \\\\
1 \times 10^{13}
\exp\left(-\frac{E_i}{k_{\mathrm{B}} T}\right), & \text{O$_2$ diffusion}.
\end{cases}
\end{equation}

Using this estimate for the evaluation-mode trajectory that dissociates 80 O atoms gives
\begin{equation}
\tau_{\mathrm{path}} = 7.50 \times 10^{-2}\ \mathrm{s} \approx 75.0\ \mathrm{ms},
\end{equation}
which corresponds to approximately $3.75 \times 10^{13}$ MD steps at a \SI{2}{fs} timestep. Even as a lower-bound estimate, this timescale is far beyond what is practical for direct MD. This highlights the key strength of the present framework: it can discover and evaluate physically plausible rare-event pathways in regimes that are effectively inaccessible to straightforward MD, while also providing the resulting evolved atomic structure after the target transformation has occurred.

\subsection{Temperature Dependence of Reaction Kinetics}
\begin{figure}[htbp]
    \centering
    \includegraphics[width=0.65\textwidth]{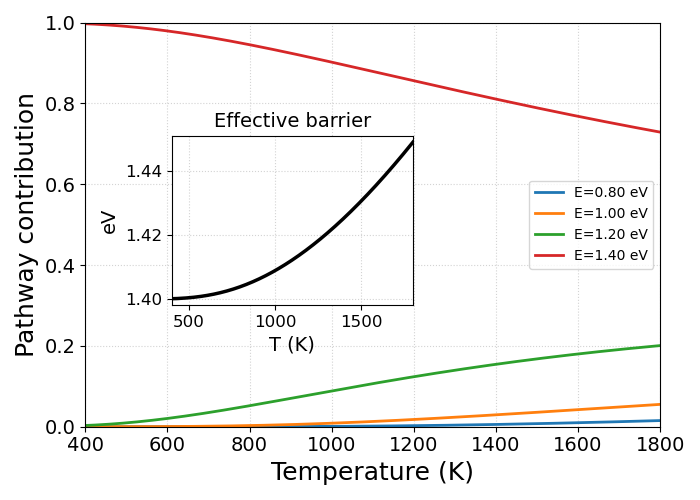}
    \caption{Temperature dependence of the barrier-dominance weights for an example barrier set with path length of 4 transitions and the corresponding effective activation barrier computed from the temperature-scaled log-sum-exp formulation.}
    \label{fig:temperature}
\end{figure}

Temperature enters the present framework through both the aggregation of pathway barriers into the effective activation barrier and the mapping of that effective barrier into the reward. Specifically, the effective barrier $E_{\mathrm{a,eff}}$ is computed using the temperature-scaled log-sum-exp formulation in  Eq. (8), and the resulting value is then converted into the activation-barrier reward $R_{\mathrm{Ea}}$ through the Fermi-Dirac-type function Eq. (9). Because both steps depend explicitly on temperature, the learned policy is formally temperature dependent. Our RL model is trained to identify sequences of atomic rearrangements that lead to dissociation while minimizing the effective activation barriers verified by NEB. When multiple activation barriers are encountered along a pathway, their combined effect is represented through the effective activation barrier defined by Eq. (8), which acts as a smooth temperature-dependent aggregation of the barrier set. Figure \ref{fig:temperature} illustrates this behavior: at low temperature the effective barrier is dominated by the largest barrier, whereas at higher temperature additional barriers contribute more appreciably to the aggregate value. Under this formulation, at low temperatures, the exponential weighting emphasizes the largest barrier, and the effective barrier approaches the largest $E_i$. In this regime, the kinetics are dominated by the rate-limiting barrier along the pathway. As temperature increases, the dominance of the single largest barrier is reduced, allowing the path length to contribute to the effective pathway difficulty. The effective barrier therefore reflects a smooth collective measure of pathway difficulty rather than a single transition state. Temperature also affects how $E_{\mathrm{a,eff}}$ is translated into reward. In the Fermi-Dirac-type form of $R_{\mathrm{Ea}}$ (Eq. (9)), temperature sets the thermal energy scale through $k_{\mathrm{B}}T$, which controls how sharply the reward changes around the reference barrier $\mu = \SI{1.2}{eV}$. At lower temperature, the smaller value of $k_{\mathrm{B}}T$ makes the reward more sensitive to small differences in $E_{\mathrm{a,eff}}$ near $\mu$, leading to a sharper distinction between favorable and unfavorable pathways. At higher temperature, the larger thermal energy broadens this transition, so the reward varies more gradually with $E_{\mathrm{a,eff}}$. Thus, temperature influences both the effective barrier assigned to a pathway and the sensitivity with which that barrier is rewarded.

In the present work we approximate the activation barriers using zero-temperature NEB energies and neglect entropic and vibrational free-energy contributions, which are typically much smaller than the barrier differences that control pathway selection and therefore do not significantly alter the relative ordering of competing mechanisms.~\cite{fultz2010vibrational} As mentioned in Sec. 2.10.2, the temperature was set to 1000 K in this work to match typical experimental conditions for dry oxidation of silicon. At this temperature, $k_{\mathrm{B}}T \approx \SI{0.086}{eV}$, which sets the thermal scale in the reward mapping.

Within the present framework, this perspective provides a natural way to interpret the set of barriers discovered along an RL trajectory. The policy identifies kinetically plausible pathways in a heterogeneous environment, while the temperature-dependent aggregation of pathway barriers and the temperature-dependent reward mapping together determine the effective difficulty of executing that pathway. In practice, the RL-generated pathway ensemble can therefore be viewed as a catalogue of mechanistically feasible events whose relative importance depends on both temperature and the distribution of barriers along the trajectory. This formulation provides a stable and differentiable proxy for pathway difficulty while preserving the dominant influence of the largest kinetic bottlenecks.

\subsection{Generality of the Framework}
Although we focus on O$_2$ dissociation near Si/a-SiO$_2$ as a concrete benchmark, the framework is intended as a general control layer for goal-directed atomistic simulation in complex environments. Its core design choices are broadly relevant to processing and microstructure evolution at the atomic scale. These include representing reactants and products as agents, potentially in a multi-agent setting, enforcing strict $E(3)$ equivariance in state and action representations, using barrier-aware objectives for reaction and diffusion events, and applying verification and rollback mechanisms to maintain physical plausibility.

These ingredients are not specific to silicon oxidation. They can be adapted to other technologically important processes in which atomic-scale dynamics control outcomes, including battery interfaces and evolving interphases, heterogeneous catalysis and reaction networks, semiconductor processing such as atomic layer deposition, and general synthesis and structural evolution. In many of these settings, key transformations such as solid-state reactions, amorphization, nucleation at defects, and grain-boundary chemistry are governed by rare events in highly heterogeneous landscapes. By choosing appropriate agents and process-specific objectives, the same framework can be used to discover feasible pathways and generate realistic endpoint structures.

More broadly, the framework can be coupled in the future to more transferable first-principles-based machine-learning interatomic potentials and uncertainty-aware active-learning workflows, which would further extend its applicability across chemistries and processing conditions. Unlike many generative models, which can produce plausible endpoint structures without a physically grounded transformation pathway, the present framework retains explicit physical information throughout the evolution of the environment. It therefore provides physical interpretation and evaluation not only of the final product, but also of the sequence of events by which that product is reached.

\section{Conclusion}

We introduced REALIZE, an $E(3)$-equivariant deep reinforcement learning framework for atomistic long-timescale simulation in experimentally relevant environments. The central idea is to move beyond predefined reaction coordinates and representation-specific search, and instead learn a symmetry-consistent control policy that discovers kinetically favorable pathways directly in a closed-loop atomistic environment. In this way, REALIZE defines a goal-directed and barrier-aware operating mode for atomistic simulation in disordered systems where conventional rare-event methods often remain limited by prior-knowledge requirements or search inefficiency.

As a demanding benchmark, we applied the framework to silicon dry oxidation, a technologically important process governed by rare O$_2$ diffusion and dissociation events in an amorphous oxide network. In this setting, the trained policy learned to avoid obvious failure modes, increased the frequency of MD-verified dissociation events, and progressively lowered the effective activation barrier of the discovered pathways. The absolute barrier scale reached by the learned policy is also broadly consistent with first-principles studies of O$_2$ diffusion in amorphous SiO$_2$, supporting the physical plausibility of the discovered transport behavior.

The structural analysis of the rate-limiting saddles points to another important feature. The learned policy does not appear to be explained by a single obvious geometric quantity. In particular, the decrease in effective activation barrier is accompanied by a decrease in the barrier-weighted saddle-point void volume, rather than the increase that might be expected from a simple wider-bottleneck picture. Together with prior first-principles results showing that O$_2$ diffusion barriers are not controlled monotonically by a single geometric descriptor, this suggests that kinetically favorable pathways in disordered oxides are influenced by more subtle local structural features than can be captured by a simple human-selected metric. This, in turn, highlights the potential value of a learned barrier-aware search framework for such systems.

Beyond training, the learned policy can also be used in a post-training evaluation mode to generate oxidation trajectories without further PPO updates. In this mode, the framework produced continued oxidation in both the Si/a-SiO$_2$ structure used during training and an out-of-training-domain Si/$\beta$-tridymite SiO$_2$ structure. The latter result is particularly encouraging because it shows that the learned policy is not tied to a single memorized environment, but can transfer to a distinct oxide morphology and still identify physically reasonable insertion and dissociation pathways. The comparison between the amorphous and crystalline starting structures further suggests that amorphous channels provide a more favorable transport environment for oxidation than initially crystalline oxide.

The present work has several limitations that also define clear directions for future study. First, the current implementation uses ReaxFF for structural relaxation, barrier estimation, and verification dynamics. Although this choice makes long-horizon closed-loop training computationally feasible, more accurate MLIP or first-principles descriptions will be valuable for refining the quantitative barrier landscape and defect statistics. Second, the current benchmark focuses on a single O$_2$ agent and a specific oxidation objective. Extending the framework to multiple interacting agents, richer action spaces, and more complex process-level objectives will be important for broader synthesis and processing applications. Third, the learned pathways invite more detailed structural analysis, including ring statistics, coordination environments, and transition-state topology, to better understand what local features the policy is exploiting.

More broadly, REALIZE opens a route toward atomistic simulation of processing and synthesis problems in which the target transformation is known but the productive pathways are not. In that regime, the combination of explicit physical verification, kinetic preference, and symmetry-preserving action generation offers a practical route to bridge atomistic simulation with experimentally relevant processing conditions. We expect this framework to be applicable well beyond silicon oxidation, including other diffusion-limited interfacial reactions, thin-film growth, catalytic surface transformations, and materials synthesis problems in which rare events in disordered environments govern the realized structure.

\section*{Code availability}
The code used in this study will be made publicly available upon formal publication of this manuscript in a peer-reviewed journal.

\section*{Author contributions}
W.J. conceived the reinforcement-learning framework, developed and implemented the model, performed the simulations and analyses, prepared the figures, and wrote the manuscript. B.D. supervised the project, provided scientific guidance, contributed to interpretation of the results, and revised the manuscript. F.T. provided project oversight and contributed to manuscript revision.

\section*{Acknowledgements}
The authors acknowledge the National Institute of Standards and Technology for providing computational resources through the Center for Theoretical and Computational Materials Science and Redwing high-performance computing clusters. This work was performed with funding from the CHIPS Metrology Program, part of CHIPS for America, National Institute of Standards and Technology, U.S. Department of Commerce. Certain equipment, instruments, software, or materials are identified in this paper in order to specify the experimental procedure adequately. Such identification is not intended to imply recommendation or endorsement of any product or service by NIST, nor is it intended to imply that the materials or equipment identified are necessarily the best available for the purpose.

\section*{Competing interests}
The authors declare no competing interests.

\printbibliography

\end{document}